\documentclass[useAMS,usenatbib]{mn2e}
\usepackage{epsfig}
\title[Chemical composition of the Orion nebula]{A reappraisal of the chemical composition of the Orion nebula 
based on VLT echelle spectrophotometry\thanks{Based on observations collected at the European Southern
Observatory, Chile, proposal number ESO 68.C-0149(A)}}
\author[C. Esteban et al.]
  {C.~Esteban,$^1$
  M.~Peimbert,$^2$
  J.~Garc\'\i a-Rojas,$^1$ M.T.~Ruiz,$^3$
  \newauthor 
  A.~Peimbert,$^2$ and 
  M.~Rodr\'\i guez,$^4$ \\
  $^1$Instituto de Astrof\'\i sica de Canarias, E-38200 La Laguna, Tenerife, Spain \\
  $^2$Instituto de Astronom\'\i a, UNAM, Apdo. Postal 70-264, M\'exico 04510 D.F., Mexico\\
  $^3$Departamento de Astronom\'\i a, Universidad de Chile, Casilla Postal 36D, Santiago de Chile, Chile\\
  $^4$Instituto Nacional de Astrof\'\i sica, \'Optica y Electr\'onica INAOE, Apdo. Postal 51 y 216, 
  7200 Puebla, Pue., Mexico\\}

\newcommand{\te}{$T_{\rm e}$}
\newcommand{\cubiccm}{cm$^{-3}$}
\newcommand{\hydp}{H$^+$}

\newcommand{\cpp}{C$^{++}$}

\newcommand{\clp}{Cl$^+$}
\newcommand{\clpp}{Cl$^{++}$}
\newcommand{\clppp}{Cl$^{3+}$}
\newcommand{\op}{O$^+$}
\newcommand{\opp}{O$^{++}$}

\newcommand{\np}{N$^+$}
\newcommand{\npp}{N$^{++}$}

\newcommand{\nepp}{Ne$^{++}$}

\newcommand{\arpp}{Ar$^{++}$}
\newcommand{\arppp}{Ar$^{3+}$}
\newcommand{\sulp}{S$^+$}
\newcommand{\sulpp}{S$^{++}$}
\newcommand{\fep}{Fe$^{+}$}
\newcommand{\fepp}{Fe$^{++}$}
\newcommand{\feppp}{Fe$^{3+}$}
\newcommand{\Hb}{H$\beta$}

\newcommand{\fci}{[C~{\sc i}]}
\newcommand{\foiii}{[O~{\sc iii}]}

\newcommand{\foi}{[O~{\sc i}]}
\newcommand{\foii}{[O~{\sc ii}]}
\newcommand{\fsii}{[S~{\sc ii}]}
\newcommand{\fsiii}{[S~{\sc iii}]}

\newcommand{\fnii}{[N~{\sc ii}]}
\newcommand{\fariv}{[Ar~{\sc iv}]}
\newcommand{\fcliii}{[Cl~{\sc iii}]}

\newcommand{\ffeii}{[Fe~{\sc ii}]}
\newcommand{\ffeiii}{[Fe~{\sc iii}]}
\newcommand{\ffeiv}{[Fe~{\sc iv}]}
\newcommand{\oiii}{O~{\sc iii}}
\newcommand{\nitroi}{N~{\sc i}}
\newcommand{\nii}{N~{\sc ii}}
\newcommand{\niii}{N~{\sc iii}}
\newcommand{\sili}{Si~{\sc i}}
\newcommand{\silii}{Si~{\sc ii}}
\newcommand{\siliii}{Si~{\sc iii}}
\newcommand{\oi}{O~{\sc i}}
\newcommand{\oii}{O~{\sc ii}}
\newcommand{\cii}{C~{\sc ii}}
\newcommand{\nei}{Ne~{\sc i}}
\newcommand{\neii}{Ne~{\sc ii}}
\newcommand{\neiii}{Ne~{\sc iii}}
\newcommand{\sii}{S~{\sc ii}}
\newcommand{\siii}{S~{\sc iii}}

\newcommand{\niqii}{Ni~{\sc ii}}

\newcommand{\fariii}{[Ar~{\sc iii}]}
\newcommand{\fei}{Fe~{\sc i}}
\newcommand{\feii}{Fe~{\sc ii}}

\newcommand{\hi}{H\,{\sc i}}
\newcommand{\hii}{H~{\sc ii}}
\newcommand{\hei}{He~{\sc i}}

\newcommand{\hp}{H$^+$}
\newcommand{\hep}{He$^+$}

\newcommand{\tmthree}{10$^{-3}$}

\begin{document}

\maketitle

\begin{abstract}

We present VLT UVES echelle spectrophotometry of the Orion nebula in the 3100 to 10400 \AA\ range.
We have measured the intensity of 555 emission lines, many of them corresponding to permitted 
lines of different heavy-element ions. This is the largest set of spectral emission lines ever 
obtained for a Galactic or extragalactic {\hii} region. 
We have derived {\hep}, {\cpp}, {\op}, {\opp} and {\nepp} abundances from pure 
recombination lines. This is the first time that {\op} and {\nepp} abundances are obtained from this 
kind of lines in the nebula. We have also derived 
abundances from collisionally excited lines for 
a large number of ions of different elements. In all cases, ionic abundances 
obtained from recombination lines are larger than those derived from collisionally excited lines. 
We have obtained remarkably consistent independent estimations of the temperature fluctuations parameter, $t^2$, 
from different methods, which are also similar to other estimates from the literature. This result strongly suggests 
that moderate temperature fluctuations --$t^2$ between 0.02 and 0.03-- are present in the Orion nebula. 
We have compared the  
chemical composition of the nebula with those of the Sun and other representative objects. 
The heavy element abundances in the Orion nebula are only slightly higher than the solar ones, a difference that can be 
explained by the chemical evolution of the solar vicinity. 

\end{abstract}

\begin{keywords}
ISM: abundances -- {\hii} regions -- ISM: individual: Orion nebula.
\end{keywords}

\section{Introduction}

The Orion nebula is the brightest and nearest Galactic {\hii} region in the sky and 
the most observed object of this kind. Our present-day knowledge about this remarkable
nebula has been recently reviewed by \citet{ode01} and
\citet{fer01}. The chemical 
composition of the Orion nebula has been traditionally considered the standard 
reference for the ionized gas in the solar neighborhood. Much work has been devoted 
to study the chemical abundances of this object \citep[e.g.,]
[ Osterbrock, Tran \& Veilleux 1992; Esteban et al. 1998, hereinafter EPTE]{pei77,
rub91,bal91}. 

The analysis of the intensity ratios of collisionally excited lines (hereinafter CELs)
has been the usual method for determining the ionic abundances in ionized nebulae.
Peimbert, Storey \& Torres-Peimbert (1993) were the first in determining the {\opp/\hp} ratio from the intensity of
the faint {\oii} recombination lines (hereinafter RLs) in the Orion nebula. These authors find that the 
{\opp/\hp} ratio obtained from RLs is a factor of 2 
larger than that derived from CELs. The RLs of heavy element ions that can be 
detected in the optical range are very faint, of
the order of {\tmthree} or less of the intensity of H$\beta$. The brightest optical 
RLs in photoionized nebulae are those of {\cii} $\lambda$ 4267 {\AA} and the multiplet 1 of {\oii} around
$\lambda$ 4650 \AA. The difference between the abundances determined from CELs and RLs (ofted 
called {\it abundance discrepancy}) can be of the order of 5 or even 20 for some 
planetary nebulae \citep[see compilations by][]{rol94,mat99}. In the case of {\hii}
regions the discrepancy seems to be present but not as large as in the case of the extreme
planetary nebulae. \citet{EPTE,est99a,est99b} have analyzed deep echelle spectra in
several slit positions of the Orion nebula, M17 and M8, determining {\cpp} and {\opp}
abundances (as well as the {\op} abundance in the case of M8) from CELs and RLs. The
abundance discrepancies are similar for the different ions and slit positions for 
each nebula, reaching factors from 1.2 to 2.2. In more recent papers, \citet{est02}, 
\citet{pei03} and \citet{tsa03} have estimated the abundance discrepancy for several
extragalactic {\hii} regions in M33, M101 and the Magellanic Clouds finding 
discrepancies rather similar to those found in the Galactic objects. These results 
are really puzzling, because a substantial part of our knowledge about the chemical 
composition of astronomical objects --and specially those in the extragalactic 
domain-- is based on the analysis of CELs in ionized nebulae. 

One of the most probable causes of the abundance discrepancy is the presence of 
spatial variations or fluctuations in the temperature structure of the nebulae
\citep{pei67}. Recent discussions and reviews about this problem can be found in 
\citet{sta02}, \citet{liu02,liu03}, \citet{est02a}, and \citet{tor03}. The 
relation between both phenomena is possible due to the different functional 
dependence of the
line emissivities of CELs and RLs on the electron temperature, which is stronger  
--exponential-- in the case of CELs. Traditionally, following Peimbert's formalism, 
the temperature fluctuations are parametrized by $t^2$, the mean square temperature 
fluctuation of the gas. EPTE, \citet{est99a,est99b, est02} and \citet{pei03} have 
found that values of $t^2$ between 0.02 and 0.04 can account for the observed 
abundance discrepancy in the Galactic and extragalactic {\hii} regions where RLs 
have been measured. 

The main aim of this work is to make a reappraisal of the chemical composition of the
Orion nebula in one of the slit positions observed by \citet{pei77} and EPTE but including new 
echelle spectrophotometry obtained with the VLT. These new observations are described 
in the following section and give an unprecedently wider wavelength coverage for 
high-resolution spectroscopic observations of the Orion nebula. A total number of 
555 lines are detected and measured, an important improvement with respect to the 220 
lines observed by EPTE 
and the 444 ones identified --but partially analysed-- by \citet{bal00}. Abundance determinations of 
additional heavy element ions based on RLs, as {\op}, {\nepp} or {\npp} are now possible, 
as well as abundance determinations of {\opp} and {\cpp} based on additional lines not detected or  
identified in previous works.

\section[]{Observations and data reduction}

The observations were made on 2002 March 12 at Cerro Paranal Observatory 
(Chile), using the UT2 (Kueyen) of the Very Large Telescope (VLT) with the
Ultraviolet Visual Echelle Spectrograph, UVES \citep{dod00}. Two different
settings --the standard ones-- were used in both arms of the spectrograph
covering from 3100 \AA\ to 10400 \AA. Some narrow spectral ranges could not
be observed, 5783$-$5830 \AA\ and 8540$-$8650 \AA, due to the physical
separation between the two CCDs of the detector system of the red arm, and 
10084$-$10088 \AA\ and 10252$-$10259 \AA, because the last two orders of
the spectrum do not fit within the size of the CCD. 

The full width at half maximum (FWHM) of the spectral resolution at a given
wavelength is $\Delta\lambda$ $\approx$ $\lambda$/8800. The slit position was
chosen to cover approximately the same area of the Position 2 observed by  
EPTE. As in that previous work, the slit position
was oriented east-west and centred at 25 arcsec South and 10 arcsec West of
$\theta^1$Ori C, the brightest star of the Trapezium Cluster and the main
ionizing source of the Orion nebula. The atmospheric dispersor corrector (ADC)
was used during the observations to keep the same observed region within the slit
independently of the change of the parallactic angle of the object during the 
night. The slit width was set to 3.0~arcsec as a compromise between the spectral
resolution needed for the project and the desired signal-to-noise ratio of the
spectra. The slit length was fixed to 10~arcsec in the blue arm and 12~arcsec in
the red arm to avoid overlapping between consecutive orders in the spatial
direction. Five individual exposures of 60 or 120 seconds were added to obtain the
definitive spectra. Complementary shorter 5 seconds spectra were taken to obtain good
intensity measurements for the brightest emission lines, which were close to saturation 
in the longer spectra. The one-dimensional spectra were extracted for an area of 
3~$\times$~8.5 arcsec$^2$. 

The spectra were reduced using the {\sc IRAF}\footnote{IRAF is distributed by NOAO, which is
operated by AURA, under cooperative agreement with NSF} echelle reduction package
following the standard procedure of bias subtraction, aperture extraction,
flat-fielding, wavelength calibration and flux calibration. The correction for
atmospheric extinction was performed using the average curve for the continuous
atmospheric extinction at La Silla Observatory. The flux calibration was 
achieved by taking echellograms of the standard star EG~274. A journal of the 
observations is presented in Table~1. 

\setcounter{table}{0}
\begin{table}
\begin{minipage}{75mm}
\centering \caption{Journal of observations.}
\begin{tabular}{c@{\hspace{2.8mm}}c@{\hspace{2.8mm}}c@{\hspace{1.8mm}}}
\noalign{\hrule} \noalign{\vskip3pt}
Date & $\Delta\lambda~(\AA)$ & Exp. time (s) \\
\noalign{\vskip3pt} \noalign{\hrule} \noalign{\vskip3pt}
2002 March 12 & 3000$-$3900 & 5, 5$\times$60 \\
" & 3800$-$5000 & 5, 5$\times$120 \\
" & 4750$-$6800 & 5, 5$\times$60 \\
" & 6700$-$10400 & 5, 5$\times$120 \\
\noalign{\vskip3pt} \noalign{\hrule} \noalign{\vskip3pt}
\end{tabular}
\end{minipage}
\end{table}

\section[]{Line intensities and reddening}

Line intensities were measured integrating all the 
flux in the line between two given limits and over a local continuum 
estimated by eye. In the cases of evident line-blending, the line flux of each 
individual line was derived from a multiple Gaussian profile 
fit procedure. All these measurements were made with the {\sc SPLOT} routine of 
the {\sc IRAF} package. 

\setcounter{table}{1}
\begin{table}
\begin{minipage}{75mm}
\centering \caption{Observed and reddening-corrected line ratios 
[F(H$\beta$)=100] and identifications.}
\begin{tabular}{l@{\hspace{2.8mm}}l@{\hspace{2.8mm}}c@{\hspace{1.8mm}}
l@{\hspace{2.8mm}}c@{\hspace{2.8mm}}c@{\hspace{2.8mm}}r@{\hspace{1.8mm}}}
\noalign{\hrule} \noalign{\vskip3pt}
$\lambda_{\rm 0}$& & & $\lambda_{\rm obs}$& & & err \\
($\AA$)& Ion& Mult.& ($\AA$)& $F(\lambda)$& $I(\lambda)$& 
(\%) \\
\noalign{\vskip3pt} \noalign{\hrule} \noalign{\vskip3pt}
3187.84&     He I &      3&  3187.92&   1.691&   2.796 &    8\\
3276.04&     C II &       &  3276.20&   0.064&   0.102 &   : \\
3296.77&     He I &      9&  3296.93&   0.085&   0.135 &   30\\
3322.54& [Fe III] ? &    5F &  3322.68&   0.044&   0.069 &   31\\
3323.75&    Ne II &      7&  3323.87&   0.037&   0.058 &   36\\
3324.87&    S III &      2&  3325.01&   0.047&   0.074 &   29\\
3334.87&    Ne II &      2&  3334.97&   0.060&   0.094 &   24\\
3354.42&     He I &      8&  3354.72&   0.135&   0.210 &   13\\
3367.05&    Ne II &     12&  3367.30&   0.034&   0.054 &   37\\
3367.22&    Ne II &     19&         &        &         &     \\
3387.13&    S III &      2&  3387.27&   0.078&   0.120 &   20\\
3388.46&    Ne II &     19&  3388.57&   0.020&   0.030 &   : \\
3447.59&     He I &      7&  3447.76&   0.219&   0.332 &    9\\
3450.39&  [Fe II] &   27F &  3450.49&   0.027&   0.041 &   : \\
3453.07&    Ne II &     21&  3453.51&   0.015&   0.023 &   : \\
       &        ? &       &  3454.82&   0.013&   0.020 &   : \\
3456.83&     N II &       &  3457.07&   0.025&   0.038 &   : \\
3461.01&  Ca I] ? &       &  3461.17&   0.027&   0.041 &   : \\
3465.94&     He I &       &  3466.12&   0.024&   0.036 &   : \\
3471.80&     He I &       &  3471.97&   0.042&   0.063 &   30\\
3478.97&     He I &     48&  3479.14&   0.041&   0.062 &   25\\
3487.73&     He I &     42&  3487.91&   0.058&   0.087 &   25\\
3498.66&     He I &     40&  3498.84&   0.075&   0.112 &   20\\
3511.10&      O I &       &  3511.30&   0.017&   0.025 &   : \\
3512.52&     He I &     38&  3512.69&   0.092&   0.137 &   17\\
3530.50&     He I &     36&  3530.68&   0.128&   0.189 &   18\\
3536.80&     He I &       &  3536.93&   0.010&   0.015 &   : \\
3536.81&     He I &       &         &        &         &     \\
3536.93&     He I &       &         &        &         &     \\
3554.42&     He I &     34&  3554.62&   0.162&   0.237 &   11\\
3587.28&     He I &     32&  3587.47&   0.234&   0.340 &    9\\
3613.64&     He I &      6&  3613.82&   0.342&   0.493 &    7\\
3631.95& [Fe III] ? &       &  3632.16&   0.025&   0.036 &   : \\
3634.25&     He I &     28&  3634.43&   0.346&   0.495 &    7\\
3651.97&     He I &     27&  3652.16&   0.017&   0.024 &   : \\
3661.22&      H I &   H31 &  3661.41&   0.204&   0.290 &    9\\
3662.26&      H I &   H30 &  3662.43&   0.250&   0.355 &    8\\
3663.40&      H I &   H29 &  3663.59&   0.236&   0.335 &    8\\
3664.68&      H I &   H28 &  3664.86&   0.247&   0.350 &    9\\
3666.10&      H I &   H27 &  3666.29&   0.292&   0.414 &    7\\
3667.68&      H I &   H26 &  3667.87&   0.336&   0.475 &    7\\
3669.47&      H I &   H25 &  3669.66&   0.375&   0.531 &    6\\
3671.48&      H I &   H24 &  3671.67&   0.412&   0.583 &    6\\
3673.76&      H I &   H23 &  3673.95&   0.447&   0.632 &    6\\
3676.37&      H I &   H22 &  3676.56&   0.519&   0.733 &    6\\
3679.36&      H I &   H21 &  3679.55&   0.588&   0.830 &    6\\
3682.81&      H I &   H20 &  3683.00&   0.644&   0.908 &    5\\
3686.83&      H I &   H19 &  3687.02&   0.684&   0.962 &    5\\
3691.56&      H I &   H18 &  3691.75&   0.802&   1.127 &    4\\
3694.22&    Ne II &      1&  3694.39&   0.030&   0.042 &   30\\
3697.15&      H I &   H17 &  3697.34&   0.960&   1.347 &    4\\
3703.86&      H I &   H16 &  3704.04&   1.090&   1.527 &    4\\
3705.04&     He I &     25&  3705.20&   0.513&   0.717 &    5\\
3709.37&    S III &      1&  3709.67&   0.035&   0.048 &   : \\
3711.97&      H I &   H15 &  3712.16&   1.303&   1.820 &    4\\
3712.74&     O II &      3&  3712.85&   0.025&   0.035 &   : \\
3713.08&    Ne II &      5&  3713.23&   0.033&   0.046 &   : \\
3717.72&    S III &      6&  3717.92&   0.059&   0.083 &   24\\
3721.83&  [S III] &    2F &  3722.04&   2.481&   3.453 &    4\\
3721.94&      H I &   H14 &         &        &         &     \\
\noalign{\vskip3pt} \noalign{\hrule} \noalign{\vskip3pt}
\end{tabular}
\end{minipage}
\end{table}

\setcounter{table}{1}
\begin{table}
\begin{minipage}{75mm}
\centering \caption{{\it --continued}}
\begin{tabular}{l@{\hspace{2.8mm}}l@{\hspace{2.8mm}}c@{\hspace{1.8mm}}
l@{\hspace{2.8mm}}c@{\hspace{2.8mm}}c@{\hspace{2.8mm}}r@{\hspace{1.8mm}}}
\noalign{\hrule} \noalign{\vskip3pt}
$\lambda_{\rm 0}$& & & $\lambda_{\rm obs}$& & & err \\
($\AA$)& Ion& Mult.& ($\AA$)& $F(\lambda)$& $I(\lambda)$& 
(\%) \\
\noalign{\vskip3pt} \noalign{\hrule} \noalign{\vskip3pt}
3726.03&   [O II] &    1F &  3726.30&  40.122&  55.776 &    4\\
       &       ? &       &  3727.40&   0.055&   0.076 &   : \\
3728.82&   [O II] &    1F &  3729.04&  19.366&  26.898 &    4\\
3732.86&     He I &     24&  3733.06&   0.037&   0.052 &   : \\
3734.37&      H I &   H13 &  3734.56&   1.929&   2.675 &    4\\
3737.55&    Ne II &       &  3737.85&   0.018&   0.025 &   : \\
3749.48&     O II &      3&  3749.62&   0.083&   0.115 &   18\\
3750.15&      H I &   H12 &  3750.34&   2.377&   3.280 &    4\\
3756.10&     He I &       &  3756.32&   0.043&   0.060 &   31\\
3768.78&     He I &       &  3768.99&   0.015&   0.020 &   : \\
       &     ? &       &  3769.95&   0.017&   0.023 &   : \\
3770.63&      H I &   H11 &  3770.82&   3.058&   4.193 &    4\\
3784.89&     He I &     64&  3785.07&   0.027&   0.036 &   : \\
3786.72&  [Cr II] & &  3786.90&   0.011&   0.016 &   : \\
3787.40&     He I &       &  3787.61&   0.006&   0.009 &   : \\
3797.63&  [S III] &    2F &  3798.10&   3.969&   5.394 &    3\\
3797.90&      H I &   H10 &         &        &         &     \\
3805.74&     He I &     58&  3805.96&   0.041&   0.055 &   22\\
3806.54&   Si III &      5&  3806.68&   0.017&   0.023 &   30\\
3819.61&     He I &     22&  3819.82&   0.899&   1.213 &    3\\
3829.77&    Ne II &     39&  3829.92&   0.013&   0.018 &   : \\
3831.66&     S II &       &  3831.87&   0.038&   0.051 &   12\\
3833.57&     He I &       &  3833.73&   0.043&   0.058 &   11\\
3835.39&      H I &    H9 &  3835.58&   5.407&   7.264 &    3\\
3837.73&    S III &      5&  3837.91&   0.022&   0.029 &   18\\
3838.09&     He I &     61&  3838.47&   0.048&   0.064 &   10\\
3838.37&     N II &     30&         &        &         &     \\
3853.66&    Si II &      1&  3853.90&   0.021&   0.029 &   : \\
3856.02&    Si II &      1&  3856.27&   0.146&   0.195 &    6\\
3856.13&     O II &     12&         &        &         &     \\
3860.64&     S II &     50&  3860.81&   0.019&   0.026 &   19\\
3862.59&    Si II &      1&  3862.83&   0.076&   0.102 &    9\\
3864.12&     O II &     11&  3864.54&   0.021&   0.027 &   : \\
3867.49&     He I &     20&  3867.69&   0.060&   0.080 &    9\\
3868.75& [Ne III] &    1F &  3868.94&  17.203&  22.870 &    3\\
3871.82&     He I &     60&  3871.97&   0.067&   0.089 &    8\\
3878.18&     He I &       &  3878.39&   0.012&   0.016 &   : \\
3882.19&     O II &     12&  3882.41&   0.016&   0.021 &   : \\
3888.65&     He I &      2&  3889.18&  11.380&  15.032 &    3\\
3889.05&      H I &    H8 &         &        &         &     \\
3918.98&     C II &      4&  3919.12&   0.052&   0.068 &   10\\
3920.68&     C II &      4&  3920.83&   0.109&   0.143 &    6\\
3926.53&     He I &     58&  3926.75&   0.095&   0.124 &    7\\
3928.55&    S III &       &  3928.74&   0.017&   0.022 &   18\\
3935.94&     He I &     57&  3936.18&   0.017&   0.022 &   : \\
3954.36&     O II &      6&  3954.72&   0.019&   0.025 &   : \\
3964.73&     He I &      5&  3964.93&   0.740&   0.954 &    3\\
3967.46& [Ne III] &    1F &  3967.64&   5.314&   6.849 &    3\\
3970.07&      H I &    H7 &  3970.27&  12.366&  15.925 &    3\\
3973.24&     O II &      6&  3973.45&   0.016&   0.020 &   35\\
3983.72&    S III &      8&  3983.97&   0.032&   0.040 &   15\\
3985.93&    S III &      8&  3986.12&   0.021&   0.027 &   18\\
3993.06&  [Ni II] &  &  3993.46&   0.013&   0.017 &   25\\
3994.99&     N II &     12&  3995.18&   0.008&   0.010 &   : \\
4004.15&  Fe II ? &       &  4004.24&   0.024&   0.031 &   : \\
4008.36& [Fe III] &    4F &  4008.57&   0.017&   0.022 &   21\\
4009.22&     He I &     55&  4009.46&   0.134&   0.171 &    5\\
4023.98&     He I &     54&  4024.19&   0.017&   0.021 &   22\\
4026.08&     N II &     40&  4026.41&   1.722&   2.181 &    3\\
4026.21&     He I &     18&         &        &         &     \\
\noalign{\vskip3pt} \noalign{\hrule} \noalign{\vskip3pt}
\end{tabular}
\end{minipage}
\end{table}

\setcounter{table}{1}
\begin{table}
\begin{minipage}{75mm}
\centering \caption{{\it --continued}}
\begin{tabular}{l@{\hspace{2.8mm}}l@{\hspace{2.8mm}}c@{\hspace{1.8mm}}
l@{\hspace{2.8mm}}c@{\hspace{2.8mm}}c@{\hspace{2.8mm}}r@{\hspace{1.8mm}}}
\noalign{\hrule} \noalign{\vskip3pt}
$\lambda_{\rm 0}$& & & $\lambda_{\rm obs}$& & & err \\
($\AA$)& Ion& Mult.& ($\AA$)& $F(\lambda)$& $I(\lambda)$& 
(\%) \\
\noalign{\vskip3pt} \noalign{\hrule} \noalign{\vskip3pt}
       &       ? &       &  4027.42&   0.025&   0.031 &   16\\
4041.31&     N II &     39&  4041.49&   0.010&   0.013 &   : \\
4060.60&     O II &     97&  4060.80&   0.003&   0.004 &   : \\
4062.94&     O II &     50&  4063.18&   0.005&   0.006 &   : \\
4068.60&   [S II] &    1F &  4068.92&   1.112&   1.392 &    3\\
4069.62&     O II &     10&  4069.98&   0.069&   0.086 &    8\\
4069.89&     O II &     10&         &        &         &     \\
4072.15&     O II &     10&  4072.34&   0.054&   0.067 &    9\\
4075.86&     O II &     10&  4076.06&   0.063&   0.079 &    8\\
4076.35&   [S II] &    1F &  4076.67&   0.372&   0.464 &    3\\
4078.84&     O II &     10&  4079.05&   0.009&   0.011 &   : \\
4083.90&     O II &     47&  4084.07&   0.008&   0.010 &   37\\
4085.11&     O II &     10&  4085.32&   0.011&   0.013 &   30\\
4087.15&     O II &     48&  4087.36&   0.010&   0.013 &   31\\
4089.29&     O II &     48&  4089.49&   0.020&   0.025 &   19\\
4092.93&     O II &     10&  4093.11&   0.008&   0.010 &   : \\
4095.64&     O II &     48&  4095.82&   0.005&   0.007 &   : \\
4097.22&     O II &     20&  4097.47&   0.038&   0.047 &   10\\
4097.26&     O II &     48&         &        &         &     \\
4101.74&      H I &    H6 &  4101.95&  20.231&  25.090 &    2\\
4104.99&     O II &     20&  4105.12&   0.019&   0.024 &   19\\
4107.09&     O II & 48.01 &  4107.25&   0.004&   0.006 &   : \\
4110.79&     O II &     20&  4110.94&   0.019&   0.024 &   19\\
4112.10&     Ne I &       &  4112.25&   0.006&   0.008 &   : \\
4114.48&  [Fe II] &   23F &  4114.78&   0.005&   0.006 &   : \\
4116.07& Fe II] ? &       &  4116.22&   0.006&   0.007 &   : \\
4119.22&     O II &     20&  4119.41&   0.025&   0.031 &   16\\
4120.82&     He I &     16&  4121.01&   0.179&   0.221 &    4\\
4121.46&     O II &     19&  4121.63&   0.033&   0.041 &   13\\
4129.32&     O II &     19&  4129.48&   0.006&   0.008 &   : \\
4131.89& [Fe III] &       &  4131.94&   0.013&   0.016 &   30\\
4132.80&     O II &     19&  4132.98&   0.027&   0.033 &   15\\
4143.76&     He I &     53&  4143.96&   0.233&   0.285 &    4\\
4145.90&    O II  &    106&  4146.31&   0.011&   0.014 &   29\\
4146.08&    O II  &    106&         &        &         &     \\
4153.30&     O II &     19&  4153.47&   0.062&   0.076 &    8\\
4156.36&     N II &     19&  4156.53&   0.059&   0.072 &    9\\
4168.97&     He I &     52&  4169.28&   0.049&   0.060 &   10\\
4185.45&     O II &     36&  4185.65&   0.017&   0.021 &   21\\
4189.79&     O II &     36&  4189.96&   0.021&   0.025 &   18\\
4201.35&     N II &     49&  4201.59&   0.005&   0.006 &   : \\
4219.76&    Ne II &     52&  4219.92&   0.007&   0.008 &   : \\
4236.91&     N II &     48&  4237.25&   0.006&   0.007 &   : \\
4237.05&     N II &     48&         &        &         &     \\
4241.78&     N II &     48&  4241.97&   0.010&   0.012 &   : \\
4242.49&     N II &     48&  4242.80&   0.010&   0.012 &   : \\
4243.97&  [Fe II] &   21F &  4244.37&   0.035&   0.042 &   12\\
4249.08&  [Fe II] &  &  4249.25&   0.006&   0.008 &   : \\
4253.54&    S III &      4&  4253.79&   0.035&   0.041 &   13\\
4267.15&     C II &      6&  4267.38&   0.201&   0.238 &    4\\
4275.55&     O II &     67&  4275.76&   0.014&   0.017 &   24\\
4276.75&     O II &     67&  4277.20&   0.027&   0.032 &   15\\
4276.83&  [Fe II] &   21F &         &        &         &     \\
4287.39&  [Fe II] &    7F &  4287.79&   0.065&   0.087 &    8\\
4294.78&     S II &     49&  4294.83&   0.015&   0.018 &   23\\
4294.92&    O II  &     54&         &        &         &     \\  
4300.66&  Fe II ? &       &  4300.81&   0.055&   0.065 &    9\\
4303.82&     O II &     53&  4304.02&   0.014&   0.017 &   24\\
4303.82&     O II &     53&         &        &         &     \\
4307.23&     O II &     54&  4307.43&   0.006&   0.007 &   : \\
4317.14&     O II &      2&  4317.31&   0.038&   0.044 &   12\\
4319.63&     O II &      2&  4319.84&   0.022&   0.025 &   18\\
\noalign{\vskip3pt} \noalign{\hrule} \noalign{\vskip3pt}
\end{tabular}
\end{minipage}
\end{table}

\setcounter{table}{1}
\begin{table}
\begin{minipage}{75mm}
\centering \caption{{\it --continued}}
\begin{tabular}{l@{\hspace{2.8mm}}l@{\hspace{2.8mm}}c@{\hspace{1.8mm}}
l@{\hspace{2.8mm}}c@{\hspace{2.8mm}}c@{\hspace{2.8mm}}r@{\hspace{1.8mm}}}
\noalign{\hrule} \noalign{\vskip3pt}
$\lambda_{\rm 0}$& & & $\lambda_{\rm obs}$& & & err \\
($\AA$)& Ion& Mult.& ($\AA$)& $F(\lambda)$& $I(\lambda)$& 
(\%) \\
\noalign{\vskip3pt} \noalign{\hrule} \noalign{\vskip3pt}
4325.76&     O II &      2&  4325.95&   0.014&   0.017 &   24\\
4326.40&      O I &       &  4326.66&   0.026&   0.031 &   15\\
4326.24& [Ni II]  & 2D-4P &         &        &         &     \\
4332.69&     O II &     65&  4332.90&   0.018&   0.020 &   21\\
4336.79&  [Cr II] & a6D-a2&  4337.04&   0.019&   0.022 &   19\\
4340.47&      H I & H$\gamma$&  4340.69&  38.720&  44.932 &    2\\
4344.35&   O I] ? &       &  4344.53&   0.005&   0.006 &   : \\
4345.55&     O II & 63.01 &  4345.72&   0.055&   0.064 &    9\\
4345.56&     O II &      2&         &        &         &     \\
4346.85&  [Fe II] &   21F &  4347.42&   0.013&   0.015 &   : \\
4349.43&     O II &      2&  4349.62&   0.056&   0.065 &    9\\
4351.26&     O II &     16&  4351.46&   0.007&   0.008 &   : \\
4352.78& [Fe II]  &   21F &  4353.17&   0.010&   0.012 &   25\\
4359.34&  [Fe II] &    7F &  4359.74&   0.050&   0.058 &   10\\
4361.54&    S III &      4&  4361.73&   0.014&   0.016 &   25\\
4363.21&  [O III] &    2F &  4363.42&   1.129&   1.301 &    2\\
4364.61&  Mn II ? &       &  4364.86&   0.005&   0.005 &   : \\
4366.89&     O II &      2&  4367.06&   0.042&   0.048 &   11\\
4368.19&      O I &      5&  4368.66&   0.063&   0.073 &    9\\
4368.25&      O I &      5&         &        &         &     \\
4375.72&     Ne I &       &  4376.12&   0.008&   0.009 &   : \\
4387.93&     He I &     51&  4388.15&   0.473&   0.542 &    2\\
4391.94&    Ne II &     57&  4392.14&   0.012&   0.014 &   27\\
4409.30&    Ne II &     57&  4409.50&   0.008&   0.009 &   36\\
4413.78&  [Fe II] &    7F &  4414.19&   0.036&   0.036 &   13\\
4414.90&     O II &      5&  4415.09&   0.032&   0.036 &   16\\
4416.27&  [Fe II] &    6F &  4416.67&   0.040&   0.045 &   14\\
4416.97&     O II &      5&  4417.16&   0.024&   0.028 &   16\\
4422.36&  Ni II ? &       &  4422.51&   0.005&   0.005 &   : \\
4422.37&  Cr II ? &       &         &        &         &     \\
4428.54&    Ne II &     57&  4428.71&   0.008&   0.009 &   : \\
4432.51&     Ne I &       &  4432.76&   0.009&   0.010 &   : \\
4432.54&     Ne I &       &         &        &         &     \\
4437.55&     He I &     50&  4437.78&   0.063&   0.071 &    8\\
4452.11&  [Fe II] &    7F &  4452.51&   0.029&   0.033 &   14\\
4452.38&     O II &      5&         &        &         &     \\
4457.95&  [Fe II] &    6F &  4458.37&   0.017&   0.020 &   21\\
4465.41&     O II &     94&  4465.67&   0.015&   0.017 &   23\\
4467.92&     O II &     94&  4468.15&   0.008&   0.009 &   : \\
4471.09&     He I &     14&  4471.72&   4.042&   4.523 &    1\\
4474.91&  [Fe II] &    7F &  4475.32&   0.012&   0.013 &   28\\
4491.14&  [Fe IV] &       &  4491.45&   0.009&   0.010 &   33\\
4492.64&  [Fe II] &    6F &  4493.07&   0.009&   0.010 &   34\\
4514.90&  [Fe II] &    6F &  4515.26&   0.007&   0.008 &   : \\
4571.20&    Mg I] &      1&  4571.44&   0.005&   0.005 &   : \\
4590.97&     O II &     15&  4591.18&   0.023&   0.025 &   17\\
4592.43&    Fe I ? &       &  4592.62&   0.005&   0.005 &   : \\
4595.95&     O II &     15&  4596.38&   0.019&   0.020 &   20\\
4596.18&     O II &     15&         &        &         &     \\
4596.83& [Ni III] &  &  4597.26&   0.005&   0.005 &   : \\
4601.48&     N II &      5&  4601.69&   0.012&   0.013 &   27\\
4602.11&     O II &     93&  4602.34&   0.005&   0.006 &   : \\
4607.16&     N II &      5&  4607.37&   0.039&   0.042 &   12\\
4607.13& [Fe III] &    3F &         &        &         &     \\
4609.44&     O II &     93&  4609.68&   0.012&   0.013 &   27\\
4613.87&     N II &      5&  4614.07&   0.010&   0.010 &   32\\
4620.11&   C II ? &       &  4620.83&   0.015&   0.016 &   24\\
4620.26&   C II ? &       &         &        &         &     \\
4621.39&     N II &      5&  4621.62&   0.015&   0.016 &   24\\
4628.05&  [Ni II] &  &  4628.49&   0.006&   0.007 &   : \\
4630.54&     N II &      5&  4630.76&   0.044&   0.048 &   10\\
\noalign{\vskip3pt} \noalign{\hrule} \noalign{\vskip3pt}
\end{tabular}
\end{minipage}
\end{table}

\setcounter{table}{1}
\begin{table}
\begin{minipage}{75mm}
\centering \caption{{\it --continued}}
\begin{tabular}{l@{\hspace{2.8mm}}l@{\hspace{2.8mm}}c@{\hspace{1.8mm}}
l@{\hspace{2.8mm}}c@{\hspace{2.8mm}}c@{\hspace{2.8mm}}r@{\hspace{1.8mm}}}
\noalign{\hrule} \noalign{\vskip3pt}
$\lambda_{\rm 0}$& & & $\lambda_{\rm obs}$& & & err \\
($\AA$)& Ion& Mult.& ($\AA$)& $F(\lambda)$& $I(\lambda)$& 
(\%) \\
\noalign{\vskip3pt} \noalign{\hrule} \noalign{\vskip3pt}
4634.14&    N III &      2&  4634.31&   0.016&   0.018 &   22\\
4638.86&     O II &      1&  4639.05&   0.053&   0.057 &    9\\
4640.64&    N III &      2&  4640.80&   0.027&   0.029 &   13\\
4641.81&     O II &      1&  4642.02&   0.096&   0.102 &    5\\
4641.85&    N III &      2&         &        &         &     \\
4643.06&     N II &      5&  4643.31&   0.014&   0.015 &   25\\
4649.13&     O II &      1&  4649.35&   0.146&   0.155 &    3\\
4650.84&     O II &      1&  4651.04&   0.049&   0.052 &   10\\
4658.10& [Fe III] &    3F &  4658.42&   0.517&   0.549 &    2\\
4661.63&     O II &      1&  4661.81&   0.064&   0.068 &    8\\
4667.01& [Fe III] &    3F &  4667.25&   0.029&   0.031 &   14\\
4673.73&     O II &      1&  4673.99&   0.011&   0.011 &   29\\
4676.24&     O II &      1&  4676.43&   0.033&   0.035 &   13\\
4696.36&     O II &      1&  4696.60&   0.004&   0.004 &   : \\
4699.22&     O II &     25&  4699.39&   0.010&   0.010 &   32\\
4701.62& [Fe III] &    3F &  4701.88&   0.165&   0.172 &    4\\
4705.35&     O II &     25&  4705.57&   0.018&   0.018 &   21\\
4710.07&     Ne I &     11&  4710.23&   0.007&   0.007 &   : \\
4711.37&  [Ar IV] &    1F &  4711.56&   0.096&   0.100 &    6\\
4713.14&     He I &     12&  4713.41&   0.657&   0.685 &    1\\
4728.07&  [Fe II] &    4F &  4728.45&   0.005&   0.005 &   : \\
4733.93& [Fe III] &    3F &  4734.20&   0.066&   0.069 &    8\\
4740.16&  [Ar IV] &    1F &  4740.42&   0.116&   0.121 &    5\\
4752.95&     O II &       &  4753.15&   0.010&   0.010 &   31\\
4754.83& [Fe III] &    3F &  4755.05&   0.100&   0.103 &    6\\
 4769.6& [Fe III] &    3F &  4769.77&   0.060&   0.061 &    8\\
4772.18&  Cr II ? &       &  4772.46&   0.005&   0.006 &   : \\
4774.74&  [Fe II] &   20F &  4775.16&   0.009&   0.010 &   33\\
4777.88& [Fe III] &    3F &  4778.02&   0.032&   0.033 &   11\\
4779.71&     N II &     20&  4779.99&   0.011&   0.011 &   29\\
4788.13&     N II &     20&  4788.37&   0.014&   0.014 &   25\\
4802.36& [Co II] ? &       &  4802.75&   0.011&   0.011 &   29\\
4803.29&     N II &     20&  4803.55&   0.018&   0.019 &   20\\
4814.55&  [Fe II] &   20F &  4815.00&   0.040&   0.041 &   11\\
4815.51&     S II &      9&  4815.84&   0.016&   0.016 &   22\\
4861.33&      H I & H$\beta$ &  4861.61& 100.000& 100.000 & 0.7\\
4881.00& [Fe III] &    2F &  4881.40&   0.255&   0.254 &    3\\
4889.70&  [Fe II] &       &  4890.11&   0.026&   0.026 &   15\\
4890.86&     O II &     28&  4891.09&   0.022&   0.022 &   19\\
4895.05&      N I &     78&  4895.21&   0.015&   0.015 &   24\\
4902.65&    Si II &   7.23&  4902.91&   0.014&   0.013 &   25\\
4905.34&  [Fe II] &    20F&  4905.88&   0.016&   0.015 &   23\\
4921.93&     He I &     48&  4922.23&   1.240&   1.222 &    1\\
4924.50& [Fe III] &    2F &  4924.76&   0.050&   0.049 &   10\\
4924.53&     O II &     28&         &        &         &     \\
4930.50& [Fe III] &    1F &  4930.98&   0.021&   0.021 &   18\\
4931.32&  [O III] &    1F &  4931.53&   0.053&   0.052 &    9\\
4943.04&     O II &     33&  4943.41&   0.010&   0.010 &   : \\
4947.38&  [Fe II] &   20F &  4947.86&   0.008&   0.008 &   : \\
4949.39&   Ar II ? &       &  4949.54&   0.007&   0.007 &   : \\
4958.91&  [O III] &    1F &  4959.22& 131.389& 128.202 &  0.7\\
4968.63&    Cr II &       &  4968.94&   0.010&   0.010 &   : \\
4980.13&      O I &       &  4980.42&   0.013&   0.012 &   26\\
4985.90& [Fe III] &    2F &  4986.15&   0.012&   0.012 &   27\\
4987.20& [Fe III] &    2F &  4987.62&   0.047&   0.046 &   10\\
4987.38&     N II &     24&         &        &         &     \\
4994.37&     N II &     24&  4994.74&   0.018&   0.018 &   35\\
4997.02&  MnII ? &       &  4997.28&   0.036&   0.035 &   18\\
5001.13&     N II &     19&  5001.72&   0.031&   0.030 &   16\\
5001.47&     N II &     19&         &        &         &     \\
5006.84&  [O III] &    1F &  5007.19& 398.147& 383.804 &  0.7\\
5011.30& [Fe III] &    1F &  5011.72&   0.070&   0.067 &   14\\
\noalign{\vskip3pt} \noalign{\hrule} \noalign{\vskip3pt}
\end{tabular}
\end{minipage}
\end{table}

\setcounter{table}{1}
\begin{table}
\begin{minipage}{75mm}
\centering \caption{{\it --continued}}
\begin{tabular}{l@{\hspace{2.8mm}}l@{\hspace{2.8mm}}c@{\hspace{1.8mm}}
l@{\hspace{2.8mm}}c@{\hspace{2.8mm}}c@{\hspace{2.8mm}}r@{\hspace{1.8mm}}}
\noalign{\hrule} \noalign{\vskip3pt}
$\lambda_{\rm 0}$& & & $\lambda_{\rm obs}$& & & err \\
($\AA$)& Ion& Mult.& ($\AA$)& $F(\lambda)$& $I(\lambda)$& 
(\%) \\
\noalign{\vskip3pt} \noalign{\hrule} \noalign{\vskip3pt}
5015.68&     He I &      4&  5016.02&   2.397&   2.306 &    1\\
       &       ? &       &  5017.14&   0.025&   0.024 &   20\\
 5035.49& [Fe II]  &    4F &  5036.16&   0.020&   0.019 &   24\\
5041.03&    Si II &      5&  5041.40&   0.118&   0.113 &    7\\
5041.98&     O II &  23.01&  5042.32&   0.026&   0.024 &   19\\
5045.10&     N II &      4&  5045.44&   0.015&   0.014 &   20\\
5047.74&     He I &     47&  5048.33&   0.605&   0.577 &    2\\
5055.98&    Si II &      5&  5056.40&   0.207&   0.197 &    4\\
5084.77& [Fe III] &    1F &  5085.11&   0.012&   0.011 &   35\\
5111.63& [Fe II]  &   19F &  5112.25&   0.019&   0.018 &   25\\
5121.82&     C II &     12&  5122.16&   0.010&   0.009 &   : \\
 5146.61& O I &       &  5147.25&   0.040&   0.037 &   15\\
 5146.61& O I &       &           &        &         &     \\   
 5158.81& [Fe II]  &   19F &  5159.37&   0.064&   0.060 &    9\\
5191.82& [Ar III] &    3F &  5192.07&   0.072&   0.066 &    9\\
5197.90&    [N I] &    1F &  5198.50&   0.140&   0.128 &    6\\
5200.26&    [N I] &    1F &  5200.85&   0.083&   0.076 &    8\\
5219.31&    S III &       &  5219.71&   0.011&   0.010 &   38\\
5261.61& [Fe II]  &   19F &  5262.21&   0.052&   0.047 &   11\\
5270.40& [Fe III] &    1F &  5270.93&   0.305&   0.274 &    2\\
5273.38& [Fe II]  &   18F &  5273.92&   0.023&   0.021 &   21\\
5274.97&      O I &     27&  5275.69&   0.013&   0.011 &   30\\
5275.12&      O I &     27&         &        &         &     \\
5298.89&      O I &     26&  5299.60&   0.031&   0.028 &   17\\
5299.04&      O I &     26&         &        &         &     \\
5342.40&     C II &  17.06&  5342.73&   0.015&   0.013 &   30\\
5363.35&  [Ni IV] & 4F-2G &  5363.94&   0.009&   0.008 &   : \\
5405.15&    Ne II &       &  5405.30&   0.008&   0.007 &   : \\
5412.00& [Fe III] &    1F &  5412.53&   0.030&   0.026 &   17\\
5433.49&     O II &       &  5433.71&   0.008&   0.007 &   : \\
5453.81&    S II  &      6&  5454.24&   0.012&   0.010 &   : \\
5495.67&     N II &     29&  5495.98&   0.006&   0.005 &   : \\
5512.77&      O I &     25&  5513.32&   0.028&   0.024 &   18\\
5517.71&  [Cl III] &    1F &  5518.03&   0.454&   0.383 &    3\\
5537.88&  [Cl III] &    1F &  5538.20&   0.704&   0.590 &    2\\
5551.95&     N II &     63&  5552.30&   0.009&   0.007 &   : \\
5554.83&      O I &     24&  5555.55&   0.030&   0.025 &   17\\
5555.03&      O I &     24&         &        &         &     \\
5577.34&    [O I] &    3F &  5577.89&   0.010&   0.008 &   : \\
5666.64&     N II &      3&  5666.93&   0.035&   0.029 &   15\\
5676.02&     N II &      3&  5676.35&   0.012&   0.010 &   : \\
5679.56&     N II &      3&  5679.92&   0.053&   0.043 &   11\\
5686.21&     N II &      3&  5686.59&   0.008&   0.006 &   : \\
5710.76&     N II &      3&  5711.06&   0.011&   0.009 &   35\\
5739.73&   Si III &      4&  5740.05&   0.047&   0.037 &   12\\
5746.96&  [Fe II] &   34F &  5747.59&   0.006&   0.005 &   : \\
       &       ? &       &  5752.86&   0.007&   0.006 &   : \\
5754.64&   [N II] &    3F &  5755.08&   0.858&   0.680 &    3\\
5867.99&  Ni II ? &       &  5868.26&   0.026&   0.020 &   30\\
5875.64&     He I &     11&  5875.98&  18.764&  14.418 &    3\\
5906.15&   Si I ? &       &  5906.35&   0.011&   0.008 &   : \\
5927.82&     N II &     28&  5928.16&   0.013&   0.010 &   : \\
5931.78&     N II &     28&  5932.15&   0.026&   0.020 &   19\\
5941.65&     N II &     28&  5941.91&   0.020&   0.015 &   24\\
5944.38&  Fe II ? &       &  5944.70&   0.007&   0.005 &   : \\
5944.40&  Fe II ? &       &         &        &         &     \\
5952.39&     N II &     28&  5952.80&   0.017&   0.012 &   : \\
5957.56&    Si II &      4&  5958.09&   0.061&   0.046 &   10\\
5958.39&      O I &     23&  5959.19&   0.050&   0.038 &   12\\
5958.58&      O I &     23&         &        &         &     \\
5978.93&    Si II &      4&  5979.43&   0.130&   0.097 &    6\\
6000.20& [Ni III] &    2F &  6000.59&   0.015&   0.011 &   30\\
6046.23&      O I &     22&  6046.99&   0.121&   0.089 &    7\\
\noalign{\vskip3pt} \noalign{\hrule} \noalign{\vskip3pt}
\end{tabular}
\end{minipage}
\end{table}

\setcounter{table}{1}
\begin{table}
\begin{minipage}{75mm}
\centering \caption{{\it --continued}}
\begin{tabular}{l@{\hspace{2.8mm}}l@{\hspace{2.8mm}}c@{\hspace{1.8mm}}
l@{\hspace{2.8mm}}c@{\hspace{2.8mm}}c@{\hspace{2.8mm}}r@{\hspace{1.8mm}}}
\noalign{\hrule} \noalign{\vskip3pt}
$\lambda_{\rm 0}$& & & $\lambda_{\rm obs}$& & & err \\
($\AA$)& Ion& Mult.& ($\AA$)& $F(\lambda)$& $I(\lambda)$& 
(\%) \\
\noalign{\vskip3pt} \noalign{\hrule} \noalign{\vskip3pt}
6046.44&      O I &     22&         &        &         &     \\
6046.49&      O I &     22&         &        &         &     \\
6151.43&     C II &  16.04&  6151.73&   0.012&   0.009 &   36\\
6155.98&      O I &     10&  6156.27&   0.008&   0.005 &   : \\
6157.42&    Ni II &       &  6157.68&   0.008&   0.006 &   : \\
6256.83&      O I &  50.01&  6257.42&   0.016&   0.011 &   28\\
6300.30&    [O I] &    1F &  6300.91&   1.049&   0.707 &    5\\
6312.10&  [S III] &    3F &  6312.44&   2.762&   1.853 &    4\\
6347.11&    Si II &      2&  6347.55&   0.266&   0.176 &    5\\
6363.78&    [O I] &    1F &  6364.39&   0.368&   0.242 &    5\\
6365.10&  [Ni II] & 8F &  6365.72&   0.014&   0.009 &   32\\
6371.36&    Si II &      2&  6371.76&   0.149&   0.098 &    7\\
 6401.4& [Ni III] &    2F &  6401.70&   0.010&   0.007 &   : \\
6402.25&     Ne I &      1&  6402.77&   0.013&   0.009 &   : \\
6454.77&     C II &  17.05&  6455.33&   0.008&   0.005 &   : \\
6461.95&     C II &  17.04&  6462.23&   0.039&   0.025 &   15\\
 6533.8& [Ni III] &    2F &  6533.99&   0.037&   0.023 &   15\\
6548.03&   [N II] &    1F &  6548.57&  19.665&  12.201 &    5\\
6552.62&  Cr II ? &       &  6553.00&   0.024&   0.015 &   : \\
6555.84&     O II & 105.39&  6556.11&   0.012&   0.008 &   : \\
6562.82&      H I & H$\alpha$ &  6563.15& 465.402& 287.378 &    5\\
6576.48&     O II &       &  6576.71&   0.013&   0.008 &   33\\
6576.57&     O II &       &         &        &         &     \\
6578.05&     C II &      2&  6578.36&   0.473&   0.291 &    6\\
6583.41&   [N II] &    1F &  6583.94&  61.589&  37.769 &    5\\
6666.80&  [Ni II] & 8F &  6667.44&   0.024&   0.014 &   21\\
6678.15&     He I &     46&  6678.49&   6.475&   3.848 &    6\\
 6682.2& [Ni III] &    2F &  6682.23&   0.008&   0.005 &   : \\
6710.97&  [Fe II] &       &  6711.03&   0.005&   0.003 &   : \\
6716.47&   [S II] &    2F &  6716.96&   3.303&   1.938 &    6\\
6721.39&     O II &      4&  6721.71&   0.011&   0.006 &   : \\
6730.85&   [S II] &    2F &  6731.36&   6.023&   3.518 &    6\\
6734.00&     C II &    21 &  6734.42&   0.010&   0.006 &   : \\
 6739.8&  [Fe IV] &       &  6740.23&   0.009&   0.005 &   : \\
6744.39&     N II &       &  6744.42&   0.006&   0.003 &   : \\
 6747.5&  [Cr IV] ? &  &  6747.97&   0.007&   0.004 &   34\\
6755.85&     He I & 1/20&  6756.28&   0.006&   0.003 &   32\\
 6755.9&  [Fe IV] &       &         &        &         &     \\
6759.14&  [Cr II] &  &  6759.40&   0.004&   0.002 &   : \\
6760.78&  MnII ? &       &  6760.98&   0.004&   0.002 &   : \\
6769.59&      N I &     58&  6769.97&   0.009&   0.005 &   29\\
6785.81&     O II &       &  6786.05&   0.009&   0.005 &   27\\
6787.04&  Fe II ? &       &  6787.41&   0.003&   0.001 &   : \\
6791.48&  [Ni II] &     8F&  6791.97&   0.012&   0.007 &   22\\
6797.00& [Ni III] &       &  6797.12&   0.005&   0.003 &   : \\
       &        ? &       &  6809.88&   0.007&   0.004 &   34\\
6809.99&     N II &     54&  6810.46&   0.004&   0.003 &   : \\
6813.57&  [Ni II] & 8F &  6814.23&   0.008&   0.005 &   23\\
6818.42&     
Si II &       &  6818.75&   0.003&   0.002 &   : \\
6821.16& [Mn III] ? &       &  6821.68&   0.003&   0.002 &   : \\
6855.88&     He I & 1/12 &  6856.34&   0.016&   0.009 &   18\\
6933.91&     He I &       &  6934.29&   0.025&   0.014 &   14\\
6989.47&     He I &       &  6989.89&   0.024&   0.013 &   12\\
7001.92&      O I &     21&  7002.80&   0.161&   0.086 &    8\\
7002.23&      O I &     21&         &        &         &     \\
7047.13& Fe II ? &       &  7047.31&   0.010&   0.006 &   25\\
7062.26&     He I & 1/11 &  7062.65&   0.037&   0.019 &   10\\
7065.28&     He I &     10&  7065.58&  14.162&   7.398 &    7\\
7096.99&   S II ? &       &  7097.22&   0.011&   0.006 &   24\\
7097.12&     Si I &       &         &        &         &     \\
7110.90& [Cl IV]&       &  7111.12&   0.005&   0.002 &   : \\
7113.42&    Si II &   7.19&  7113.66&   0.004&   0.002 &   : \\
\noalign{\vskip3pt} \noalign{\hrule} \noalign{\vskip3pt}
\end{tabular}
\end{minipage}
\end{table}

\setcounter{table}{1}
\begin{table}
\begin{minipage}{75mm}
\centering \caption{{\it --continued}}
\begin{tabular}{l@{\hspace{2.8mm}}l@{\hspace{2.8mm}}c@{\hspace{1.8mm}}
l@{\hspace{2.8mm}}c@{\hspace{2.8mm}}c@{\hspace{2.8mm}}r@{\hspace{1.8mm}}}
\noalign{\hrule} \noalign{\vskip3pt}
$\lambda_{\rm 0}$& & & $\lambda_{\rm obs}$& & & err \\
($\AA$)& Ion& Mult.& ($\AA$)& $F(\lambda)$& $I(\lambda)$& 
(\%) \\
\noalign{\vskip3pt} \noalign{\hrule} \noalign{\vskip3pt}
7115.63&     C II &     20&  7115.92&   0.006&   0.003 &   : \\
7135.78& [Ar III] &    1F &  7136.13&  31.779&  16.197 &    7\\
7151.08&     O II & 99.01 &  7151.39&   0.006&   0.003 &   : \\
7155.14&  [Fe II] &   14F &  7155.82&   0.085&   0.043 &    9\\
7160.13&     He I & 1/10 &  7160.89&   0.055&   0.028 &   10\\
7231.34&     C II &      3&  7231.62&   0.148&   0.073 &    9\\
7236.42&     C II &      3&  7236.82&   0.494&   0.243 &    8\\
7243.99&   [Ni I] &    2F &  7244.30&   0.041&   0.020 &   12\\
7254.15&      O I &     20&  7255.06&   0.216&   0.106 &    8\\
7254.45&      O I &     20&         &        &         &     \\
7254.53&      O I &     20&         &        &         &     \\
7281.35&     He I &     45&  7281.74&   1.231&   0.597 &    8\\
7298.05&     He I & 1/9 &  7298.37&   0.077&   0.037 &   10\\
7318.39&   [O II] &    2F &  7320.45&  11.363&   5.432 &    8\\
7319.99&   [O II] &    2F &         &        &         &     \\
7329.66&   [O II] &    2F &  7330.78&   8.721&   4.154 &    8\\
7330.73&   [O II] &    2F &         &        &         &     \\
7377.83&  [Ni II] &    2F &  7378.54&   0.152&   0.071 &    9\\
7388.16&  [Fe II] &   14F &  7388.82&   0.015&   0.007 &   20\\
7411.61&  [Ni II] &    2F &  7412.34&   0.048&   0.022 &   10\\
7423.64&      N I &      3&  7424.36&   0.027&   0.012 &   15\\
7442.30&      N I &      3&  7443.04&   0.067&   0.031 &   10\\
7452.54&  [Fe II] &   14F &  7453.22&   0.033&   0.015 &   13\\
7459.30& [V II] ? &    4F &  7459.64&   0.005&   0.002 &   : \\
7468.31&      N I &      3&  7469.03&   0.096&   0.044 &   10\\
7499.85&     He I & 1/8 &  7500.21&   0.122&   0.055 &   10\\
7504.94&     O II &       &  7505.33&   0.014&   0.006 &   21\\
7519.49&     C II & 16.08 &  7520.09&   0.018&   0.008 &   18\\
7519.86&     C II & 16.08 &         &        &         &     \\
7530.57&     C II & 16.08 &  7530.76&   0.046&   0.020 &   12\\
7535.21&    N II ? &       &  7535.32&   0.008&   0.004 &   36\\
7745.10&    Si I ? &       &  7745.47&   0.008&   0.003 &   : \\
7751.10& [Ar III] &    2F &  7751.50&   8.949&   3.682 &   10\\
7771.94&      O I &      1&  7772.55&   0.040&   0.016$^{\rm a}$ &   :\\
7775.39&      O I &      1&  7775.95&   0.013&   0.006 &   21\\
7811.68&     He I &       &  7812.05&   0.009&   0.003 &   29\\
7816.13&     He I & 1/7 &  7816.52&   0.197&   0.079 &   10\\
7876.03& [P II] ? &       &  7876.59&   0.014&   0.005 &   22\\
7890.07&    Ca I] &       &  7890.50&   0.096&   0.038 &   11\\
7937.13&     He I & 4/27 &  7937.61&   0.006&   0.002 &   : \\
7971.62&     He I & 2/11&  7972.09&   0.011&   0.004 &   25\\
       &       ?  &       &  7973.58&   0.008&   0.003 &   30\\
7982.40&      O I &     19&  7982.78&   0.006&   0.002 &   : \\
7987.33&      O I &     19&  7987.82&   0.011&   0.004 &   32\\
8000.08&  [Cr II] & 1F &  8000.81&   0.029&   0.011 &   16\\
8015.67&    Ca I] &       &  8016.22&   0.005&   0.002 &   : \\
8030.65&    Ca I] &       &  8031.25&   0.011&   0.004 &   : \\
 8034.9&     Si I &       &  8035.30&   0.009&   0.003 &   : \\
8045.62&  [Cl IV] &    1F &  8046.05&   0.109&   0.041 &   12\\
   8057&     He I &  4/18 &  8057.97&   0.012&   0.005 &   24\\
   8084&     He I &  4/17 &  8084.73&   0.007&   0.002 &   : \\
8092.53&    Ca I] &       &  8092.97&   0.007&   0.002 &   : \\
8094.08&     He I &  4/10 &  8094.50&   0.014&   0.005 &   22\\
   8116&     He I &  4/16 &  8116.81&   0.015&   0.006 &   21\\
 8125.31&   Ca I] &       &  8126.02&   0.014&   0.005 &   22\\
8155.66&     He I &       &  8155.93&   0.021&   0.008 &   18\\
8200.36&      N I &      2&  8201.17&   0.027&   0.010 &   16\\
8203.85&     He I &  4/14 &  8204.31&   0.026&   0.009 &   17\\
8210.72&      N I &      2&  8211.72&   0.009&   0.003 &   29\\
8216.34&      N I &      2&  8217.02&   0.073&   0.026 &   13\\
8223.14&      N I &      2&  8223.95&   0.149&   0.053 &   12\\
8245.64&      H I &   P42 &  8246.06&   0.105&   0.037 &   12\\
\noalign{\vskip3pt} \noalign{\hrule} \noalign{\vskip3pt}
\end{tabular}
\end{minipage}
\end{table}

\setcounter{table}{1}
\begin{table}
\begin{minipage}{75mm}
\centering \caption{{\it --continued}}
\begin{tabular}{l@{\hspace{2.8mm}}l@{\hspace{2.8mm}}c@{\hspace{1.8mm}}
l@{\hspace{2.8mm}}c@{\hspace{2.8mm}}c@{\hspace{2.8mm}}r@{\hspace{1.8mm}}}
\noalign{\hrule} \noalign{\vskip3pt}
$\lambda_{\rm 0}$& & & $\lambda_{\rm obs}$& & & err \\
($\AA$)& Ion& Mult.& ($\AA$)& $F(\lambda)$& $I(\lambda)$& 
(\%) \\
\noalign{\vskip3pt} \noalign{\hrule} \noalign{\vskip3pt}
8247.73&      H I &   P41 &  8248.16&   0.117&   0.041 &   12\\
8249.20&      H I &   P40 &  8250.42&   0.125&   0.044 &   12\\
8252.40&      H I &   P39 &  8252.83&   0.129&   0.046 &   12\\
8255.02&      H I &   P38 &  8255.27&   0.076&   0.027 &   13\\
8257.85&      H I &   P37 &  8258.24&   0.137&   0.048 &   12\\
8260.93&      H I &   P36 &  8261.36&   0.173&   0.061 &   12\\
8264.28&      H I &   P35 &  8264.76&   0.207&   0.073 &   12\\
8267.94&      H I &   P34 &  8268.37&   0.182&   0.064 &   12\\
8271.93&      H I &   P33 &  8272.35&   0.199&   0.070 &   12\\
8276.31&      H I &   P32 &  8276.85&   0.268&   0.094 &   12\\
8281.12&      H I &   P31 &  8281.63&   0.181&   0.063 &   12\\
8286.43&      H I &   P30 &  8286.71&   0.161&   0.056 &   12\\
8292.31&      H I &   P29 &  8292.70&   0.272&   0.095 &   12\\
8298.83&      H I &   P28 &  8299.17&   0.261&   0.091 &   12\\
8306.11&      H I &   P27 &  8306.54&   0.336&   0.117 &   12\\
8314.26&      H I &   P26 &  8314.66&   0.368&   0.128 &   12\\
8323.42&      H I &   P25 &  8323.86&   0.435&   0.151 &   12\\
       &       ? &       &  8330.35&   0.019&   0.007 &   19\\
8333.78&      H I &   P24 &  8334.21&   0.453&   0.157 &   12\\
8342.33&     He I &  4/12 &  8342.61&   0.068&   0.023 &   13\\
8345.55&      H I &   P23 &  8345.99&   0.511&   0.176 &   12\\
8359.00&      H I &   P22 &  8359.43&   0.601&   0.207 &   12\\
8361.67&     He I &  1/6 &  8362.14&   0.336&   0.115 &   12\\
8374.48&      H I &   P21 &  8374.91&   0.636&   0.217 &   12\\
   8376&     He I &  6/20 &  8376.98&   0.021&   0.007 &   18\\
 8392.4&      H I &   P20 &  8392.84&   0.713&   0.243 &   12\\
   8397&     He I &  6/19 &  8397.68&   0.024&   0.008 &   17\\
8413.32&      H I &   P19 &  8413.79&   0.891&   0.302 &   12\\
   8422&     He I &  6/18 &  8422.41&   0.029&   0.010 &   16\\
   8424&     He I &  7/18 &  8424.66&   0.015&   0.005 &   22\\
8433.94& [Cl III] &    3F &  8434.09&   0.027&   0.009 &   17\\
8437.96&      H I &   P18 &  8438.39&   0.981&   0.330 &   12\\
8446.25&      O I &      4&  8447.28&   2.626&   0.882 &   12\\
8446.36&      O I &      4&         &        &         &     \\
8446.76&      O I &      4&         &        &         &     \\
8453.15&  Fe I] ? &       &  8453.85&   0.019&   0.006 &   19\\
8453.66&  Fe I] ? &	  &         &        &         &     \\
8459.50&    Ca I] &       &  8459.98&   0.005&   0.002 &   : \\
8467.25&      H I &   P17 &  8467.69&   1.123&   0.375 &   12\\
8476.98&  Ni II ? &       &  8477.45&   0.013&   0.004 &   : \\
8480.90& [Cl III] &    3F &  8481.28&   0.031&   0.010 &   16\\
8486.27&     He I &  6/16 &  8486.70&   0.040&   0.013 &   15\\
8488.73&     He I &  7/16 &  8489.15&   0.015&   0.005 &   22\\
8488.77&     He I &  5/16 &         &        &         &     \\
 8499.7& [Cl III]&    3F &  8500.33&   0.082&   0.027 &   13\\
8502.48&      H I &   P16 &  8502.96&   1.400&   0.463 &   12\\
8518.04&     He I &  2/8  &  8518.40&   0.030&   0.010 &   19\\
8528.99&     He I &  6/15 &  8529.44&   0.060&   0.020 &   16\\
8531.48&     He I &  7/15 &  8532.09&   0.025&   0.008 &   18\\
8665.02&      H I &   P13 &  8665.44&   2.489&   0.789 &   13\\
8680.28&      N I &      1&  8681.04&   0.105&   0.033 &   14\\
8683.40&      N I &      1&  8684.24&   0.091&   0.029 &   14\\
8686.15&      N I &      1&  8686.91&   0.078&   0.025 &   14\\
8703.25&      N I &      1&  8704.13&   0.067&   0.021 &   14\\
8711.70&      N I &      1&  8712.54&   0.069&   0.022 &   14\\
8718.83&      N I &      1&  8719.65&   0.042&   0.013 &   15\\
8727.13&    [C I] &    3F &  8727.90&   0.053&   0.017 &   15\\
8728.90& [Fe III] &    8F &  8729.83&   0.036&   0.011 &   16\\
  8728.90&     N I &      21&       &        &         &     \\
8733.43&     He I &  6/12 &  8733.87&   0.107&   0.033 &   14\\
8736.04&     He I &  7/12 &  8736.48&   0.036&   0.011 &   16\\
8739.97&     He I &  5/12 &  8740.51&   0.011&   0.003 &   27\\
8750.47&      H I &   P12 &  8750.93&   3.175&   0.985 &   13\\
\noalign{\vskip3pt} \noalign{\hrule} \noalign{\vskip3pt}
\end{tabular}
\end{minipage}
\end{table}

\setcounter{table}{1}
\begin{table}
\begin{minipage}{75mm}
\centering \caption{{\it --continued}}
\begin{tabular}{l@{\hspace{2.8mm}}l@{\hspace{2.8mm}}c@{\hspace{1.8mm}}
l@{\hspace{2.8mm}}c@{\hspace{2.8mm}}c@{\hspace{2.8mm}}r@{\hspace{1.8mm}}}
\noalign{\hrule} \noalign{\vskip3pt}
$\lambda_{\rm 0}$& & & $\lambda_{\rm obs}$& & & err \\
($\AA$)& Ion& Mult.& ($\AA$)& $F(\lambda)$& $I(\lambda)$& 
(\%) \\
\noalign{\vskip3pt} \noalign{\hrule} \noalign{\vskip3pt}
8776.77&     He I & 4/9 &  8777.39&   0.260&   0.080 &   13\\
8816.82&     He I & 10/12&  8817.08&   0.017&   0.005 &   21\\
8820.00& Fe II] ? &       &  8820.38&   0.007&   0.002 &   : \\
8829.40&  [S III] &    3F &  8830.21&   0.042&   0.013 &   16\\
8831.87&  [Cr II] & 18F &  8832.21&   0.017&   0.005 &   : \\
8838.2& [Fe III] &       &  8838.75&   0.009&   0.003 &   29\\
8845.38&     He I &  6/11 &  8845.82&   0.153&   0.046 &   14\\
8848.05&     He I &  7/11 &  8848.80&   0.108&   0.033 &   14\\
8854.11&     He I &  5/11 &  8854.51&   0.027&   0.008 &   18\\
8862.79&      H I &   P11 &  8863.24&   4.133&   1.245 &   13\\
8892.22&     Ne I &       &  8892.72&   0.035&   0.011 &   16\\
8914.77&     He I &  2/7 &  8915.18&   0.064&   0.019 &   15\\
8930.97&     He I &  10/11&  8931.16&   0.017&   0.005 &   22\\
8996.99&     He I &  6/10 &  8997.42&   0.199&   0.058 &   14\\
9014.91&      H I &   P10 &  9015.24&   3.320&   0.963 &   14\\
9015.77&   N II ? &       &  9016.42&   0.077&   0.022 &   15\\
9052.16&    Ca I] &       &  9052.85&   0.018&   0.005 &   : \\
9063.29&     He I &  4/8 &  9063.78&   0.179&   0.052 &   14\\
       &       ? &       &  9067.72&   0.031&   0.009 &   17\\
9068.90&  [S III] &    1F &  9069.42& 105.114&  30.218 &   14\\
9095.09&    Ca I] &       &  9095.94&   0.073&   0.021 &   15\\
9123.60& [Cl II]&      1F &  9124.42&   0.062&   0.018 &   15\\
9204.17&     O II &       &  9204.98&   0.044&   0.013 &   16\\
9210.28&     He I & 6/9 &  9210.79&   0.289&   0.081 &   14\\
9213.20&     He I & 7/9 &  9213.54&   0.044&   0.012 &   17\\
9218.47&    Fe I] &       &  9219.10&   0.032&   0.009 &   18\\
9229.01&      H I &    P9 &  9229.49&   7.093&   1.989 &   14\\
9463.57&     He I & 1/5 &  9464.04&   0.336&   0.091 &   15\\
9516.57&     He I & 4/7 &  9517.18&   0.110&   0.030 &   15\\
9526.16&     He I & 6/8 &  9526.66&   0.192&   0.051 &   15\\
9530.60&  [S III] &    1F &  9531.48& 271.299&  72.548 &   15\\
9535.41&     O II &       &  9536.05&   0.071&   0.019 &   16\\
9545.97&      H I &    P8 &  9546.51&   9.377&   2.502 &   15\\
9702.44& Cl I ?&       &  9702.66&   0.102&   0.027 &   16\\
9824.13&    [C I] &    1F &  9825.03&   0.061&   0.016 &   16\\
9834.7&     O II &      &  9835.46&   0.043&   0.011 &   17\\
9850.24&    [C I] &    1F &  9851.10&   0.269&   0.071 &   15\\
9903.46&     C II & 17.02 &  9904.00&   0.205&   0.052 &   16\\
9962.63&     O II & 105.06&  9963.05&   0.022&   0.005 &   : \\
10005.4&     S II &       & 10005.98&   0.047&   0.012 &   17\\
10008.6&     Ne I &       & 10009.21&   0.032&   0.008 &   19\\
10027.7&     He I & 6/7 & 10028.23&   0.784&   0.194 &   16\\
10031.2&     He I & 7/7 & 10031.65&   0.252&   0.062 &   16\\
10049.4&      H I &    P7 & 10049.91&  20.915&   5.175 &   16\\
10138.4&     He I & 10/7 & 10138.89&   0.112&   0.027 &   16\\
10286.7&   [S II] &    3F & 10287.46&   1.190&   0.288 &   16\\
10310.7&     He I & 4/6 & 10311.82&   0.538&   0.130 &   16\\
10320.5&   [S II] &    3F & 10321.24&   1.459&   0.353 &   16\\
10336.4&   [S II] &    3F & 10337.17&   1.057&   0.255 &   16\\
10344.7&      N I &       & 10345.23&   0.271&   0.065 &   16\\
10344.8&      N I &       &         &        &         &     \\
\noalign{\vskip3pt} \noalign{\hrule} \noalign{\vskip3pt}
\end{tabular}
\begin{description}
\item[$^{\rm a}$] Blend with sky emission line.
\end{description}
\end{minipage}
\end{table}

All the line intensities of a given spectrum have been normalized to a 
particular non-saturated bright emission line present in each wavelength interval. 
For the bluest spectra (3000$-$3900 \AA\ and 3800$-$5000 \AA), the reference line 
was H~9 $\lambda$ 3835 \AA. 
In the case of the spectrum covering 4750$-$6800 \AA\ the reference line was {\hei} $\lambda$ 5876 \AA. 
Finally, the reference line for the reddest spectrum (6700$-$10400 \AA) was [\sii] 
$\lambda$ 6731 \AA. 
To produce a final homogeneous set of line intensity ratios, all of them  
were re-scaled to {\Hb}. In the case of the 
bluest spectra (3000$-$3900 \AA\ and 3800$-$5000 \AA) all the intensity ratios, 
formerly referred to H9, were multiplied by the H9/{\Hb} ratio obtained in the 
short exposure spectrum of the 3800$-$5000 \AA\ 
range. The emission line ratios of the 4750$-$6800 \AA\ range were re-scaled to 
{\Hb} multiplying by the {\hei} $\lambda$ 5876 \AA/{\Hb} ratio obtained from 
the shorter exposure spectrum. In the case of the last spectral section, 
6700$-$10400 \AA, the [\sii] $\lambda$ 6731 \AA/{\Hb} ratio obtained for the 
4750$-$6800 \AA\ spectrum was the re-scaling factor used. 

The different four spectral ranges covered in the spectra have overlapping regions 
at the edges. The final intensity of a given line in the overlapping regions is the 
average of the values obtained in both spectra. The differences in the intensity 
measured for each line in overlapping spectra do not show systematic trends and 
are always of the order or smaller than the quoted line intensity uncertainties. 
The final list of observed wavelengths, identifications and line intensities 
relative to {\Hb} is presented in Table 2.  

For a given line, the observed wavelength is determined by the centre of the baseline
chosen for the flux integration procedure or the centroid of the line when a 
Gaussian fit is used (in the case of line-blending). For the lines 
measured in the overlapping spectral regions, the average of the two independent 
determinations has been adopted. The final values of the observed wavelengths are 
relative to the heliocentric reference frame. 

The identification and adopted laboratory wavelengths of the lines collected in 
Table 2 were obtained following previous identifications in the Orion nebula by 
EPTE and \citet{bal91}, the identifications for 30~Dor by \citet{pei03} and the 
compilations of \citet{moo45,moo93}, \citet{wie66} and  
The Atomic Line List v2.04\footnote{webpage at: {\tt
http://www.pa.uky.edu/{$\sim$}peter/atomic/}}. This last interactive source of nebular 
line emission data was used directly or through the {\sc EMILI}\footnote{webpage at:{\tt
http://www.pa.msu.edu/astro/software/emili/}} code \citep{sha03}. A large number 
of sky emission lines were identified --specially in the red part of the spectrum-- but 
are not included in Table 2. About 11 emission lines could not be identified in any 
of the available references. Other 34 lines show a rather dubious identification. 
In total, about 8\% of the lines are not identified or their identifications are not 
confident. The four unidentified lines reported in Table 3 of EPTE 
have been observed again and identified as faint {\cii} or {\oii} lines. 

The reddening coefficient, $C(H\beta)$, was determined by fitting iteratively the 
observed Balmer decrement to the theoretical one computed by \citet{sto95} for the 
nebular conditions determined in Section 4. Following EPTE we have used the 
reddening function, $f(\lambda)$, normalized at {\Hb} derived by \citet{cos70} for 
the Orion nebula. A linear extrapolation of this reddening function was used for 
wavelengths between 3000$-$3500 \AA. To obtain the final value of $C(H\beta)$ we have 
taken the average of the values obtained from the intensity ratios of 21 Balmer and Paschen 
lines with respect to {\Hb} $-$from H10 to P7$-$ with the exception of those 
{\hi} lines showing line blending. The final adopted value of $C(H\beta)$ is 0.76$\pm$0.08, 
which is larger than the values of 0.39$\pm$0.04 and 0.60 reported by EPTE and 
\citet{pei77} for the same zone of the nebula. Table 2 shows the reddening-corrected
line intensity ratios, $I(\lambda)/I(H\beta)$, for each line. The integrated 
reddening-corrected H$\beta$ line flux is 
9.32$\times$10$^{-11}$ erg cm$^{-2}$ s$^{-1}$. 

In the case of the Orion nebula, there are several previous works presenting 
large lists of observed emission lines (Kaler, Aller \& Bowen 1965; 
Osterbrock et al. 1992; EPTE; Baldwin et al. 2000). EPTE   
show a comparison between their datasets and those of \cite{kal65} and \cite{ost92},  
finding a good consistency with the second but detecting systematic differences with the 
older photographic data by \cite{kal65}.  We have compared our VLT line intensity ratios with 
those of the two most recent previous spectroscopic works: EPTE and \citet{bal00}. 
In Figure 1 we compare the reddening-corrected emission line ratios obtained 
in previous works and in our spectra for the lines in common by means of least-squares 
fits. The comparison with the data of EPTE 
shows a slope of 0.987, indicating a rather good consistency between both datasets.  
It must be taken into account that both observations correspond to the same zone of the 
nebula, although the integrated area is not exactly the same. On the other hand, the 
comparison with the data of \citet{bal00} gives a slope of 1.027, also fairly good, although
there is an apparent trend of a slight overestimation of the intensity of the brightest 
lines (those with log$[I(\lambda)/I(H\beta)]$ $\geq$ $-$2.5) in the dataset of \citet{bal00} 
with respect to ours. The slit position observed by \citet{bal00} does not coincide with our 
position, 
although it can be considered rather close taking into account the large angular size 
of the Orion nebula. Their position is located 25 arcsec north and 17 arcsec west of the 
centre of our slit position. We have also detected that the intensity ratios of  
the emission lines blueward of about 5000 \AA\ tend to be higher in \citet{bal00} with 
respect to the data of both EPTE and ours. This trend is not observed when the datasets 
of EPTE and ours are compared. 

In Figure 2, we show part of our flux calibrated echelle spectrum around the lines of 
multiplet 1 of {\oii}. The same spectral range is presented by EPTE and \citet{bal00}. 
Readers can compare 
the signal-to-noise ratio and the spectral resolution of each of the three sets of 
echelle spectra. 

The observational errors associated with the line intensities (in percentage of their ratio 
with respect \Hb) are also presented in Table 2. These errors include the uncertainties in the 
line intensity measurement and flux calibration as well as the propagation of the uncertainty 
in the reddening coefficient. Colons indicate errors of the order of or larger than 40\%. 

\section{Physical conditions}

The electron density, $N_e$, has been derived from the ratio of collisionally excited 
 lines of several ions and making use of {\sc NEBULAR} routines \citep{sha95} 
included in the {\sc IRAF} package. In the case of {\ffeiii}, we have obtained the value of
$N_e$ that minimizes the dispersion of the line ratios of 14 individual {\ffeiii} emission lines 
with respect to {\ffeiii} $\lambda$ 4658 \AA. The calculations for this ion have been done 
with a 34 level model-atom that uses the collision strengths of \citet{zha96} and the 
transition probabilities of \citet{qui96}.  The {\foii} electron density has been 
obtained from two different line ratios $I(3729)/I(3726)$ and 
$I(3726+3729)/I(7319+7320+7331+7332)$. The contribution of the intensities of 
the {\foii} $\lambda\lambda$ 7319, 7320, 7331, and 7332 lines due to recombination 
has been taken into account following the expresion given by \citet{liu00}. In 
any case, this contribution is rather small (about 3\% of the total intensity). 

\setcounter{table}{2}
\begin{table}
\begin{minipage}{75mm}
\centering \caption{Physical conditions.}
\begin{tabular}{lcc}
\noalign{\hrule} \noalign{\vskip3pt}
Parameter & Line & Value \\
\noalign{\vskip3pt} \noalign{\hrule} \noalign{\vskip3pt}
$N_e$ ({\cubiccm}) & [{\nitroi}] & 1700$\pm$600 \\
 & {\foii}$^{\rm a}$ & 2400$\pm$300 \\
 & {\foii}$^{\rm b}$ & 6650$\pm$400 \\
 & {\fsii} & 6500$^{+2000}_{-1200}$ \\
 & {\ffeiii} & 9800$\pm$300 \\
  & {\fcliii} & 9400$^{+1200}_{-700}$ \\
 & {\fariv} & 6800$^{+1100}_{-1000}$ \\
 $T_e$ (K) & {\foi} & 8000: \\
 & {\fci} & $>$10000 \\
 & {\fnii} & 10150$\pm$350 \\
 & {\foii} & 9800$\pm$800 \\
 & {\fsii} & 9050$\pm$800 \\
 & {\foiii} & 8300$\pm$40 \\
 & {\fsiii} & 10400$^{+800}_{-1200}$ \\ 
 & {\fariii} & 8300$\pm$400 \\ 
 & Bac & 7900$\pm$600 \\
 & Pac & 8100$\pm$1400 \\
\noalign{\vskip3pt} \noalign{\hrule} \noalign{\vskip3pt}
\end{tabular}
\begin{description}
\item[$^{\rm a}$] From 3726/3729 ratio.
\item[$^{\rm b}$] From (3727+9)/(7319+20+31+32) ratio.
\end{description}\end{minipage}
\end{table}

From Table 3, one can see that the density obtained from the 
{\foii} $I(3729)/I(3726)$ line ratio is lower than the values obtained from most of the 
other indicators. This effect is also reported in other objects recently studied
by our group: NGC~3576 \citep{gar04} and NGC~5315 \citep{pei04} as well as marginally in  
low-density {\hii} regions as 30 Dor \citep{pei03} and NGC~2467 (Garc\'\i a-Rojas et al., in preparation), where 
$N_e$({\oii}) is somewhat lower than the densities derived from the other density indicators. 
Moreover, in the case of our data for the Orion nebula, adopting the density derived 
from {\foii} $I(3729)/I(3726)$, we find: a) the electron temperature for {\op} 
--$T_e$({\oii})-- is higher than the rest of ionic temperatures; b) a larger dispersion 
in the ionic abundances obtained from the individual [{\oii}] lines. 
Alternatively, we have derived the electron density from the 
{\foii} $I(3726+3729)/I(7319+7320+7331+7332)$ line ratio, finding that: a) the density 
is now more consistent with the rest of the indicators; b) the dispersion 
of the O$^+$/H$^+$ ratios obtained from the different individual lines is lower. Therefore, 
it seems more advisable to rely in the $N_e$({\oii}) obtained 
from the {\foii} $I(3726+3729)/I(7319+7320+7331+7332)$ ratio. We find that this indicator is also 
more consistent in the cases of NGC~3576, NGC~5315 and NGC~2467. For comparison, we have determined 
$N_e$({\oii}) from $I(3729)/I(3726)$ line ratio making use of the old {\sc FIVEL} program described 
by De Robertis, Dufour \& Hunt (1987) --the program in which {\sc NEBULAR} is based-- finding that the value obtained is higher 
(4800 cm$^{-3}$ instead of 2400 cm$^{-3}$)  
becoming more similar to those obtained from the other density indicators. We also obtain 
systematically higher --and more consistent-- values of $N_e$({\oii}) using {\sc FIVEL} for 
NGC~3576, NGC~5315, NGC~2467 and 30~Dor. The structure of both programs --{\sc FIVEL} and {\sc NEBULAR}-- is basically the same. 
Apparently, the only substantial difference is the atomic data used. {\sc NEBULAR} is periodically updated and our 
version of {\sc FIVEL} is not updated since 1996. In the case of {\oii}, {\sc FIVEL} uses the transition probabilities of 
\citet{zei82} and collision strengths of \citet{pra76} and the last version of {\sc NEBULAR} uses the 
transition probabilities recommended by \citet{wie96} and the collision strenghts of \citet{mcl93}. 
We think that the problem with the density derived from {\foii} $I(3729)/I(3726)$ ratio could be due to 
errors or problems in the atomic data used for those transitions in the latest version of {\sc NEBULAR}.

From Table 3, it seems that there are no apparent differences between densities for low and 
high-ionization-potential ions. Therefore, a value of 8900$\pm$200 {\cubiccm} has been 
adopted as representative 
of our observed zone and all ions. This is a weighted average of the densities obtained from 
the {\foii} $I(3726+3729)/I(7319+7320+7331+7332)$, {\fsii}, {\ffeiii}, {\fcliii}, 
and {\fariv} emission line ratios. This value is somewhat larger than the electron 
density of 5700 {\cubiccm} adopted by EPTE. 

As in the case of densities, electron temperatures, {\te}, have been derived from the 
ratio of collisionally excited emission lines of several ions and making use of 
{\sc NEBULAR} routines. In the case of the {\fnii} $\lambda$ 5755 {\AA} line, we have 
corrected its intensity for the contribution of recombination following \citet{liu00}. 
This contribution is very small, about 2\%.

The echelle spectra show enough good signal-to-noise ratio for the nebular 
continuum emission to allow a satisfactory determination of both the Balmer 
and Paschen discontinuities (see Figure 3). They are defined as 
$I_c(Bac)=I_c(\lambda3646^-)-I_c(\lambda3646^+)$ and 
$I_c(Pac)=I_c(\lambda8203^-)-I_c(\lambda8203^+)$ respectively. The high spectral 
resolution of the spectra permits to measure the continuum emission in zones very near de 
discontinuity, minimizing the possible contamination of other continuum 
contributions. We have obtained power-law fits to the relation between 
$I_c(Bac)/I(Hn)$ or $I_c(Pac)/I(Pn)$ and {\te} for different $n$ corresponding to 
different observed lines of both series. The emissivities as a function of electron 
temperature for the nebular continuum and the {\hi} Balmer and Paschen lines 
have been taken from \citet{bro70} and \citet{sto95} respectively. The $T_e(Bac)$ 
adopted is the average of the values using the lines from $H\alpha$ to 
H~10 (the brightest ones). In the case of $T_e(Pac)$, the adopted value is the average 
of the individual temperatures obtained using the lines from P~7 to P~18 (the brightest 
lines of the series), excluding P~8 and P~10 because their intensity seems to be 
affected by sky absorption. As it can be seen in Table 3, $T_e(Bac)$ and $T_e(Pac)$ 
are remarkably similar despite their relatively large uncertainties. 

We have adopted the average of electron temperatures obtained from {\fnii}, {\fsii}, and 
{\foii} lines 
as representative for the low ionization zone, $T_{low}$ = 10000$\pm$400 K, and the 
average of the values obtained from {\foiii}, {\fsiii}, and {\fariii} lines 
for the high ionization zone, 
$T_{high}$ = 8320$\pm$40 K. The temperatures adopted by 
EPTE were $T_{low}$ = 10710$\pm$450 K and $T_{high}$ = 8350$\pm$200 K.

\section{He$^+$ abundance}

\setcounter{table}{3}
\begin{table}
\begin{minipage}{75mm}
\centering \caption{He$^+$ abundance.}
\begin{tabular}{lcc}
\noalign{\hrule} \noalign{\vskip3pt}
Line & He$^+$/H$^+$ $^{\rm a}$\\
\noalign{\vskip3pt} \noalign{\hrule} \noalign{\vskip3pt}
3819.61 & 911 $\pm$ 27\\
3888.65 & 860 $\pm$ 26\\
3964.73 & 868 $\pm$ 26\\
4026.21 & 914 $\pm$ 27\\
4387.93 & 861 $\pm$ 17\\
4471.09 & 852 $\pm$ 9\\
4713.14 & 884 $\pm$ 9\\
4921.93 & 886 $\pm$ 9\\
5875.64 & 907 $\pm$ 27\\
6678.15 & 912 $\pm$ 55\\
7065.28 & 626 $\pm$ 44\\
7281.35 & 738 $\pm$ 59\\
Adopted & 874 $\pm$6$^{\rm b}$ \\
\noalign{\vskip3pt} \noalign{\hrule} \noalign{\vskip3pt}
\end{tabular}
\begin{description}
\item[$^{\rm a}$] In units of 10$^{-4}$, for $\tau_{3889} = 16.7 \pm 0.5$ and $t^2 = 0.022 \pm 0.002$. 
Uncertainties correspond to line intensity errors.
\item[$^{\rm b}$] It includes all the relevant uncertainties in emission line 
intensities, $N_e$, $\tau_{3889}$ and $t^2$.
\end{description}
\end{minipage}
\end{table}

We have observed a large number of {\hei} lines in our spectra. These lines arise mainly from 
recombination but they can be affected by collisional excitation and self-absorption effects. 
We have determined the {\hep}/{\hp} ratio using the effective recombination coefficients of \citet{sto95} 
for {\hi}, and those by \citet{smi96} and \citet{ben99} for {\hei}. The
collisional contribution was estimated from \citet{saw93} and \citet{kin95}, and the 
optical depth effects in the triplet lines were estimated from the computations by \citet{ben02}.
From a maximum likelihood method \citep[e. g.][]{pei00}, using $N_e = 8900 \pm 200$ cm$^{-3}$ and
$T$({\oii}+{\sc iii}) = 8730$\pm$320 K (see Sect. 8), we obtained {\hep}/{\hp} = 0.0874 $\pm$
0.0006, $\tau_{3889} = 16.7 \pm 0.5$, and $t^2 = 0.022 \pm 0.002$. In Table 4 we include the {\hep}/{\hp} 
ratios we obtain for the best observed individual {\hei} lines (those lines not affected by line blending and 
with the highest S/N for which we expect to have the best atomic data, i.e. low $n$ upper level) as well as the final adopted value, 
all the values are computed for our finally adopted $t^2 = 0.022 \pm 0.002$ (see Sect. 8). We have also 
excluded {\hei} 5015 \AA\ because it could suffer self-absorption effects from the 2$^1S$ metastable level. 
If we make a simple  ${\chi}^2$ optimisation of the values given in the table, we obtain a ${\chi}^2$ 
parameter of about 45, which indicates that the goodness of fit is rather poor. The value of 
$\tau_{3889}$=16.7 we obtain is very large and therefore the self-absorption corrections for 
triplets are large and perhaps rather uncertain. Moreover, the slit position observed is very near 
the Trapezium stars and underlying absorption by the dust-scattered stellar continua can be affecting 
the intensity of the {\hei} lines. Therefore, the adopted He$^+$ abundance can be affected by additional 
systematic uncertainties very difficult to estimate. 

\section{ionic abundances from collisionally excited lines}

\setcounter{table}{4}
\begin{table}
\begin{minipage}{75mm}
\centering \caption{Ionic abundances from collisionally excited lines$^{\rm a}$.}
\begin{tabular}{lcc}
\noalign{\hrule} \noalign{\vskip3pt}
Ion & $t^2$=0.000 & $t^2$=0.022$\pm$0.002 \\
\noalign{\vskip3pt} \noalign{\hrule} \noalign{\vskip3pt}
He$^+$ & 10.940$\pm$0.003 & 10.937$\pm$0.003 \\
N$^+$ & 6.90$\pm$0.09 & 6.96$\pm$0.09\\
O$^+$ & 7.76$\pm$0.15 & 7.90$\pm$0.15 \\
O$^{++}$ & 8.43$\pm$0.01 & 8.59$\pm$0.03 \\
Ne$^{++}$ & 7.69$\pm$0.07 & 7.86$\pm$0.07 \\
S$^+$ & 5.40$\pm$0.06 & 5.47$\pm$0.06 \\
S$^{++}$ & 7.01$\pm$0.04 & 7.18$\pm$0.05 \\
Cl$^+$ & 4.84$\pm$0.11 & 4.90$\pm$0.11 \\
Cl$^{++}$ & 5.14$\pm$0.02 & 5.30$\pm$0.02 \\
Cl$^{3+}$ & 3.79$\pm$0.12 & 3.92$\pm$0.12 \\
Ar$^{++}$ & 6.37$\pm$0.05 & 6.50$\pm$0.05 \\
Ar$^{3+}$ & 4.60$\pm$0.03 & 4.76$\pm$0.04 \\
Fe$^{++}$ & 5.37$\pm$0.08 & 5.53$\pm$0.08 \\
Fe$^{3+}$ & 5.65$^{+0.19}_{-0.30}$ & 5.78$^{+0.19}_{-0.30}$\\
\noalign{\vskip3pt} \noalign{\hrule} \noalign{\vskip3pt}
\end{tabular}
\begin{description}
\item[$^{\rm a}$] In units of 12+log(X$^m$/H$^+$).
\end{description}
\end{minipage}
\end{table}

Ionic abundances of {\np}, {\op}, {\opp}, {\nepp}, {\sulp}, {\sulpp},
{\clpp}, {\clppp}, {\arpp}, and {\arppp} have been obtained 
from collisonally excited lines (CELs) using the {\sc NEBULAR} routines of 
the {\sc IRAF} package. We have assumed a two-zone scheme and $t^2$=0, adopting 
the values of $T_{low}$ = 10000$\pm$400 K for low-ionization-potential ions
({\np}, {\op}, {\sulp}, and {\clp}) and $T_{high}$ = 8320$\pm$40 K for the 
high-ionization-potential ions ({\opp}, {\nepp}, {\sulpp}, {\clpp}, {\clppp}, {\arpp}, 
and {\arppp}). The density assumed is the same for all ions, $N_e$ = 8900$\pm$200. 
The ionic abundances are listed in Table 5.  
Many {\ffeii} lines have been identified in our spectra but all of them 
are affected by fluorescence effects \citep{rod99, ver00}.  
Unfortunately, we can not measure the {\ffeii} $\lambda$ 
8617 \AA\ line, which is almost insensitive to the effects of UV pumping. This line is precisely 
in one of the observational gaps of our spectroscopic configuration. Therefore, 
it was not possible to derive a confident value of the {\fep}/{\hydp} ratio. 
The {\fepp}/{\hydp} ratio has 
been derived from the average of the values obtained from 14 individual emission lines. 
The calculations for this ion have been done with a 34 level model-atom that uses the 
collision strengths of \citet{zha96} and the transition probabilities of \citet{qui96}. 
In the case of {\feppp}/{\hydp} ratios, we have used a 33-level model-atom where all 
collision strengths are those calculated by \citet{zha97}, the transition probabilities 
are those recommended by \citet{fro98} (and those from Garstang 1998 for the transitions 
not considered by Froese Fischer \& Rubin). The {\clp}/{\hydp} ratio cannot be derived from the {\sc NEBULAR} routines,
instead we have used an old version of the five-level atom program of \citet{sha95} 
--{\sc FIVEL}-- that is described by \citet{rob87}. This program uses the atomic data for {\clp} 
compiled by \citet{men83}. In any case, the atomic data for this ion --and therefore 
the {\clp}/{\hydp} ratio-- are rather uncertain (Shaw 2003, personal 
communication).

\section{ionic abundances of heavy elements from recombination lines}

The large sensitivity and spectral coverage of these new observations have increased 
dramatically the number of permitted lines measured in this particular zone of 
the Orion nebula with respect to the previous results of EPTE. We have detected lines 
of: {\cii}, {\nitroi}, {\nii}, {\niii}, {\oi}, 
{\oii}, {\oiii}, {\nei}, {\neii}, {\neiii}, {\sili}, {\silii}, {\siliii}, {\sii}, {\siii}, 
and perhaps some possible lines of {Mg~{\sc i}}, {Al~{\sc ii}}, {Ar~{\sc ii}}, {Cr~{\sc ii}}, 
{Mn~{\sc ii}}, {\fei}, {\feii}, and {\niqii}. 

The excitation mechanisms of many permitted lines observed in the Orion nebula have been 
discussed by Grandi (1975a, b; 1976) and EPTE. Most of these lines are 
produced by continuum and/or line fluorescence but some of them by recombination. Recombination 
lines are the only ones useful for abundance determinations. We have derived the ionic abundances 
for those ions with effective recombination coefficients available in the literature. 
EPTE only derive the {\cpp}/{\hydp} and {\opp}/{\hydp} ratios from their data but we can now also obtain 
values for {\op}/{\hydp}, {\npp}/{\hydp}, and {\nepp}/{\hydp} from recombination lines. We have also 
derived the abundances from {\nitroi} lines, but they are found to be useless because they are largely 
produced by starlight excitation. The ionic abundances obtained from permitted lines of heavy elements  
are shown in Tables 6 to 11. We have derived the abundance of the whole multiplet in the case of those 
multiplets with more than two lines observed ("Sum" in the tables). To derive the sum value we 
have used the effective recombination coefficient of the multiplet and the expected intensity 
of the whole multiplet. This last quantity has been obtained adding the intensity of the observed lines 
multiplied by the quotient of the $gf$ value of the whole multiplet with respect to the sum of 
the $gf$ values of the observed individual lines. EPTE describe the method with more detail. We prefer 
the sum value because it provides a weighted average of the abundances derived from each line of the 
multiplet and it washes out possible departures from the LTE predictions inside the multiplet. 
We have adopted $T_{high}$ for {\cpp}, {\opp}, {\npp}, and {\nepp}; $T_{low}$ for {\op} and {\np}. 
 
We have effective recombination coefficients for multiplets 
2, 3, 6, 16.04, 17.02, 17.04 and 17.06 of {\cii} \citep{dav00}. The {\cpp}/{\hydp} ratios obtained 
are shown in Table 6. The upper level of multiplet 3 
can be populated by resonance fluorescence by starlight from the ground state and this can explain 
its corresponding abnormally large {\cpp}/{\hydp} ratio. Resonance fluorescence by starlight can be also 
operating on multiplet 2 (EPTE). The rest of the multiplets included in Table 6 are produced by 
transitions involving levels with large $l$ quantum numbers and cannot be excited by permitted resonance 
transitions from the ground level. Therefore, their excitation mechanism should be recombination and 
their {\cpp}/{\hydp} ratios should reflect the true abundance of that ion. The {\cpp}/{\hydp} ratios 
obtained from the different {\cii} lines coming from large $l$ levels show an excellent agreement.  
These values are also case-independent. The final adopted {\cpp}/{\hydp} ratio is 
(22$\pm$1)$\times$10$^{-5}$. This value has been obtained from the weighted mean of the individual 
abundances obtained from multiplets 6, 16.04, 17.02, 17.04, and 17.06. In Figure 4 we show some of 
these pure recombination {\cii} lines used to derive the final {\cpp} abundance. 
EPTE obtained a 
{\cpp}/{\hydp}=20$\times$10$^{-5}$ for the same zone using the older effective recombination coefficents 
by P\'equignot, Petitjean \& Boisson (1991). All the individual abundance values used to derive the adopted average are indicated 
in boldface in Table 6. 

\setcounter{table}{5}
\begin{table*}
\centering
\begin{minipage}{150mm}
\caption{C$^{++}$/H$^+$ ratios from Permitted Lines}
\begin{tabular}{cccccc}
\noalign{\hrule} \noalign{\vskip3pt}
& & & $I$($\lambda$)/$I$(H$\beta$) & \multicolumn{2}{c}{C$^{++}$/H$^+$ ($\times$10$^{-5}$)$^{\rm a}$} \\
Mult. & Transition & $\lambda_0$ & ($\times$10$^{-2}$) & A & B \\
\noalign{\vskip3pt} \noalign{\hrule} \noalign{\vskip3pt}
2 & 3s$^2$S$-$3p$^2$P$^0$ & 6578.05 & 0.29$\pm$0.02 & 330$\pm$20 & 56$\pm$3 \\
3 & 3p$^2$P$^0$$-$3d$^2$D & 7231.34 & 0.073$\pm$0.007 & 1900$\pm$200 & 2700$\pm$300 \\
  & & 7236.42 & 0.24$\pm$0.02 & 3700$\pm$700 & 5200$\pm$400 \\
  & & Sum & 0.54$\pm$0.04 & 3700$\pm$300 & 4300$\pm$300 \\
6 & 3d$^2$D$-$4f$^2$F$^0$ & 4267.26 & 0.24$\pm$0.01 & {\bf 22$\pm$1} & $-$ \\
16.04 & 4d$^2$D$-$6f$^2$F$^0$ & 6151.43 & 0.009$\pm$0.003 & {\bf 20$\pm$7} & $-$ \\
17.02 & 4f$^2$F$^0$$-$5g$^2$G & 9903.46 & 0.052$\pm$0.008 & {\bf 19$\pm$3} & $-$ \\
17.04 & 4f$^2$F$^0$$-$6g$^2$G & 6461.95 & 0.025$\pm$0.004 & {\bf 21$\pm$3} & $-$ \\
17.06 & 4f$^2$F$^0$$-$7g$^2$G & 5342.40 & 0.013$\pm$0.004 & {\bf 23$\pm$7} & $-$ \\
\noalign{\vskip3pt} \noalign{\hrule} \noalign{\vskip3pt}
\multicolumn{3}{c}{Adopted} & & \multicolumn{2}{c}{\bf 22$\pm$1} \\
\noalign{\vskip3pt} \noalign{\hrule}
\end{tabular}
\begin{description}
\item[$^{\rm a}$] Effective recombination coefficients by Davey et al. (2000).
\end{description}
\end{minipage}
\end{table*}

Grandi (1975a) showed that the upper levels of the transitions of multiplets 1, 2, and 3 of 
{\nitroi} should be significantly populated by starlight excitation. In Table 7, we show the {\np}/{\hydp} 
ratios we obtain using the effective recombination coefficients of \citet{peq91}. The abnormally 
large abundances obtained indicate that starlight excitation is the dominant mechanism of those 
multiplets, therefore the abundances derived from the observed {\nitroi} are --unfortunately-- 
useless for our purposes and will not be considered. 

\setcounter{table}{6}
\begin{table*}
\centering
\begin{minipage}{150mm}
\caption{N$^+$/H$^+$ ratios from Permitted Lines}
\begin{tabular}{cccccc}
\noalign{\hrule} \noalign{\vskip3pt}
& & & $I$($\lambda$)/$I$(H$\beta$) & \multicolumn{2}{c}{N$^+$/H$^+$ ($\times$10$^{-5}$)$^{\rm a}$} \\
Mult. & Transition & $\lambda_0$ & ($\times$10$^{-2}$) & A & B \\
\noalign{\vskip3pt} \noalign{\hrule} \noalign{\vskip3pt}
1 & 3s$^4$P$-$3p$^4$D$^0$ & 8680.28 & 0.033$\pm$0.005 & 95$\pm$13 & 92$\pm$13 \\
  & & 8683.40 & 0.029$\pm$0.004 & 160$\pm$20 & 150$\pm$20 \\
  & & 8686.15 & 0.025$\pm$0.004 & 350$\pm$50 & 340$\pm$50 \\
  & & 8703.25 & 0.021$\pm$0.003 & 270$\pm$40 & 260$\pm$40 \\  
  & & 8711.70 & 0.022$\pm$0.003 & 240$\pm$40 & 230$\pm$40 \\
  & & 8718.83 & 0.013$\pm$0.002 & 180$\pm$30 & 180$\pm$30 \\ 
  & & Sum & 0.15$\pm$0.02 & 170$\pm$20 & 160$\pm$20 \\
2 & 3s$^4$P$-$3p$^4$P$^0$ & 8210.72 & 0.003$\pm$0.001 & 120$\pm$40 & 110$\pm$40 \\
  & & 8216.34 & 0.026$\pm$0.003 & 160$\pm$20 & 140$\pm$20 \\
  & & 8223.14 & 0.053$\pm$0.006 & 780$\pm$90 & 670$\pm$80 \\
  & & Sum & 0.15$\pm$0.02 & 330$\pm$40 & 280$\pm$30 \\
3 & 3s$^4$P$-$3p$^4$S$^0$ & 7423.64 & 0.012$\pm$0.002 & 1200$\pm$200 & 390$\pm$60 \\
  & & 7442.30 & 0.031$\pm$0.003 & 1500$\pm$200 & 490$\pm$50 \\
  & & 7468.31 & 0.044$\pm$0.004 & 1400$\pm$100 & 460$\pm$50 \\
  & & Sum & 0.09$\pm$0.01 & 1400$\pm$200 & 460$\pm$50 \\
\noalign{\vskip3pt} \noalign{\hrule} \noalign{\vskip3pt}
\end{tabular}
\begin{description}
\item[$^{\rm a}$] Effective recombination coefficients by P\'equignot et al. (1991).
\end{description}
\end{minipage}
\end{table*}

We have measured a large number of {\nii} lines in our spectra. 
\citet{gra76} showed that multiplets 3 and 5 of {\nii} in the Orion nebula may be excited by resonance fluorescence 
via the {\hei} $\lambda$ 508.6 {\AA} line. \citet{tsa03} also suggest that {\nii} triplet lines of  
the spectra of their sample {\hii} regions can be affected by fluorescence. The ground state of {\nii} is a triplet 
and, therefore, singlet lines are expected to be produced by pure recombination and should not be affected by 
fluorescence effects. We have only poor detections of three very weak singlet lines, which are not confident 
for abundance determinations. Moreover, the 
brightest singlet line reported could be a misidentification. There are three different sets of effective recombination 
coefficients available for {\nii} 
\citep{esc90,peq91,kis02}, the {\npp}/{\hydp} ratios obtained for all the lines and sets of coefficients are 
shown in Table 8. We have adopted case B as representative for triplets and obtained quite similar values of the 
{\npp}/{\hydp} ratio for all the triplet multiplets observed. We have obtained a weighted mean 
of the abundance considering multiplets 3, 4, 5, 11 and 22 (sum values of the multiplet when more than two lines 
of the multiplet are reported) and the effective recombination coefficients 
of \citet{esc90} and multiplets 3, 12, 24 and 28 and the coefficients of \citet{peq91}, finding the same 
value in both cases: {\npp}/{\hydp}=12$\times$10$^{-5}$. This value is somewhat lower than the final adopted 
abundance using the most recent effective recombination coefficients by \citet{kis02} and the weighted 
mean of the {\npp}/{\hydp} ratios obtained using multiplets 3, 4, 5, 19, 20, 24 and 28. In fact, from Table 8, 
it is clear that the individual values of the abundance obtained using \citet{kis02} are always 
somewhat larger 
than those obtained with the other two sources of effective recombination coefficients. 
All the individual abundance values used to derive the adopted average are indicated 
in boldface in Table 8. This final {\npp}/{\hydp} ratio gives a total N abundance which is 
abnormally high (see Sect. 9) independently of the recombination coefficients set used, 
indicating that the lines used in Table 8 for deriving the abundance are not produced 
by pure recombination and, unfortunately, not suitable for abundance determinations. This result has been also 
obtained by \citet{tsa03}. 

\setcounter{table}{7}
\begin{table*}
\centering
\begin{minipage}{150mm}
\caption{N$^{++}$/H$^+$ ratios from Permitted Lines}
\begin{tabular}{cccccccccc}
\noalign{\hrule} \noalign{\vskip3pt}
& & & & \multicolumn{6}{c}{N$^{++}$/H$^+$ ($\times$10$^{-5}$)} \\
& & & $I$($\lambda$)/$I$(H$\beta$) & \multicolumn{2}{c}{E\&V90$^{\rm a}$} & 
\multicolumn{2}{c}{PPB91$^{\rm b}$} & \multicolumn{2}{c}{K\&S02$^{\rm c}$} \\
Mult. & Transition & $\lambda_0$ & ($\times$10$^{-2}$)& A & B & A & B & A & B \\
\noalign{\vskip3pt} \noalign{\hrule} \noalign{\vskip3pt}
1 & 2p$^3$$^1$D$^0$$-$3p$^1$P & 4895.11 & 0.015$\pm$0.004 & 36$\pm$9 & $-$& $-$& $-$& $-$& $-$ \\
3 & 3s$^3$P$^0$$-$3p$^3$D & 5666.64 & 0.029$\pm$0.004 & 9$\pm$1 & 8$\pm$1 & 12$\pm$2 & 10$\pm$2 & 16$\pm$2 & 13$\pm$2 \\
  & & 5676.02 & 0.010: & 7: & 6: & 9: & 8: & 13: & 10: \\
  & & 5679.56 & 0.043$\pm$0.004 & 7$\pm$1 & 6$\pm$1 & 10$\pm$1 & 8$\pm$1 & 13$\pm$1 & 11$\pm$1 \\
  & & 5686.21 & 0.006: & 6: & 5: & 8: & 6: & 10: & 8: \\  
  & & 5710.70 & 0.009$\pm$0.003 & 9$\pm$3 & 8$\pm$3 & 11$\pm$4 & 9$\pm$3 & 15$\pm$5 & 13$\pm$4 \\
  & & Sum & 0.112$\pm$0.009 & 8$\pm$1 & 7$\pm$1 & 10$\pm$1 & 8$\pm$1 & 14$\pm$1 & {\bf 11$\pm$1}  \\
4 & 3s$^3$P$^0$$-$3p$^3$S & 5045.10 & 0.014$\pm$0.003 & 70$\pm$20 & 12$\pm$2 & $-$& $-$& 170$\pm$40 & 
{\bf 23$\pm$5} \\
5 & 3s$^3$P$^0$$-$3p$^3$P & 4601.48 & 0.013$\pm$0.004 & 60$\pm$20 & 11$\pm$3 & $-$& $-$& 100$\pm$30 & 17$\pm$5 \\
  & & 4613.87 & 0.010$\pm$0.003 & 100$\pm$30 & 19$\pm$6 & $-$& $-$& 170$\pm$60 & 30$\pm$7 \\
  & & 4621.39 & 0.016$\pm$0.004 & 110$\pm$30 & 20$\pm$5 & $-$& $-$& 180$\pm$40 & 32$\pm$3 \\
  & & 4630.54 & 0.048$\pm$0.005 & 70$\pm$7 & 13$\pm$1 & $-$& $-$& 110$\pm$10 & 20$\pm$2 \\
  & & 4643.06 & 0.015$\pm$0.004 & 65$\pm$10 & 12$\pm$2 & $-$& $-$& 110$\pm$20 & 19$\pm$3 \\
  & & Sum & 0.115$\pm$0.006 & 73$\pm$4 & 14$\pm$1 & $-$& $-$& 120$\pm$6 & {\bf 21$\pm$1} \\
12 & 3s$^1$P$^0$$-$3p$^1$D & 3994.99 & 0.010: & 13: & 12: & $-$& $-$& 11: & 11: \\
19 & 3p$^3$D$-$3d$^3$F$^0$ & 5001.47 & 0.030$\pm$0.005 & $-$& $-$& 9$\pm$1 & 9$\pm$1 & 7$\pm$1 & 
{\bf 7$\pm$1} \\
20 & 3p$^3$D$-$3d$^3$D$^0$ & 4803.29 & 0.019$\pm$0.004 & 9$\pm$2 & 9$\pm$2 & $-$& $-$& 12$\pm$2 & 24$\pm$4 \\
   & & 4779.71 & 0.011$\pm$0.003 & 14$\pm$4 & 14$\pm$4 & $-$& $-$& 19$\pm$6 & 40$\pm$10 \\
   & & 4788.13 & 0.014$\pm$0.004 & 12$\pm$3 & 11$\pm$3 & $-$& $-$& 16$\pm$4 & 31$\pm$8 \\
   & & Sum & 0.056$\pm$0.006 & 11$\pm$1 & 11$\pm$1 & $-$& $-$& 15$\pm$2 & {\bf 28$\pm$3} \\  
24 & 3p$^3$S$-$3d$^3$P$^0$ & 4994.37 & 0.018$\pm$0.006 & 23$\pm$8 & 22$\pm$8 & $-$ & 18$\pm$6 & 
700$\pm$200 & {\bf 30$\pm$10} \\
28 & 3p$^3$P$-$3d$^3$D$^0$ & 5927.82 & 0.010: & $-$& $-$& $-$& 25: & 1800: & 35: \\
   & & 5931.78 & 0.020$\pm$0.004 & $-$& $-$& $-$& 21$\pm$4 & 1600$\pm$300 & 30$\pm$6 \\
   & & 5941.65 & 0.015$\pm$0.004 & $-$& $-$& $-$& 9$\pm$2 & 600$\pm$200 & 12$\pm$3 \\
   & & 5952.39 & 0.012: & $-$& $-$& $-$& 39: & 2800: & 55: \\
   & & Sum & 0.063$\pm$0.005 & $-$& $-$& $-$& 17$\pm$1 & 1200$\pm$100 & {\bf 24$\pm$2} \\
29 & 3p$^1$S$-$5d$^1$P$^0$ & 5495.70 & 0.005: & $-$& $-$& $-$& 4: & $-$& $-$ \\
39 & 3d$^3$F$^0$$-$4f$^\prime$[3$\frac{1}{2}$] & 4041.31 & 0.013: &    $-$& $-$& $-$& 3: & $-$& $-$ \\
\noalign{\vskip3pt} \noalign{\hrule} \noalign{\vskip3pt}
\multicolumn{3}{c}{Adopted} & & & & & & \multicolumn{2}{c}{\bf 20$\pm$1} \\
\noalign{\vskip3pt} \noalign{\hrule}
\end{tabular}
\begin{description}
\item[$^{\rm a}$] Effective recombination coefficients by Escalante \& Victor (1990).
\item[$^{\rm b}$] Effective recombination coefficients by P\'equignot et al. (1991).
\item[$^{\rm c}$] Effective recombination coefficients by Kisielius \& Storey (2002).
\end{description}
\end{minipage}
\end{table*}

Several {\oi} lines are identified and measured in our spectra. Most of them correspond to transitions 
between triplet levels that can be excited from the ground state (2$p^4$ $^3P$) by starlight excitation, as it 
was demostrated by \citet{gra75b}. We have measured lines of multiplet 1 of {\oi}, which 
corresponds to transition between quintet levels. In principle, these lines should be produced by pure 
recombination and are also case-insensitive. Lines of multiplet 1 of {\oi} are in a spectral 
region with numerous sky emission lines. Unfortunately, the combination of our spectral resolution and the radial velocity 
of Orion nebula does not permit to deblend the brightest line of multiplet 1 at $\lambda$ 7771.94 \AA\ and an underlying 
sky emission feature. Therefore, we have to rely on the {\op}/{\hydp} ratio obtained from the faint 
{\oi} $\lambda$ 7775.34 \AA\ line, which has a large uncertainty. In any case, this is the first time the  
{\op} abundance is derived from RLs in the Orion nebula. We have two sets of effective recombination coefficents available 
for {\oi} in the literature, those by \citet{esc92} and \citet{peq91}, both sets give quite similar values of the abundances. 
In Table 9, we show the {\op}/{\hydp} ratios obtained for the different 
useful lines and multiplets. The values obtained from triplet lines are always much larger than those 
obtained from multiplet 1, demostrating the important contribution of starlight excitation to the intensity 
of the triplet lines.  

\setcounter{table}{8}
\begin{table*}
\centering
\begin{minipage}{150mm}
\caption{O$^+$/H$^+$ ratios from Permitted Lines}
\begin{tabular}{cccccccc}
\noalign{\hrule} \noalign{\vskip3pt}
& & & & \multicolumn{4}{c}{O$^+$/H$^+$ ($\times$10$^{-5}$)} \\
& & & $I$($\lambda$)/$I$(H$\beta$) &  \multicolumn{2}{c}{E\&V92$^{\rm a}$} & 
\multicolumn{2}{c}{PPB91$^{\rm b}$}  \\
Mult. & Transition & $\lambda_0$ & ($\times$10$^{-2}$)& A & B & A & B \\
\noalign{\vskip3pt} \noalign{\hrule} \noalign{\vskip3pt}
1 & 3s$^5$S$^0$$-$3p$^5$P & 7771.94 & 0.016$^{\rm c}$ & 21: & $-$ & 16: & $-$ \\
  & & 7775.34 & 0.006$\pm$0.001 & {\bf 16$\pm$3} & $-$ & {\bf 12$\pm$2} & $-$ \\
4 & 3s$^3$S$^0$$-$3p$^3$P & 8446.48 & 0.9$\pm$0.1 & 5100$\pm$600 & 1000$\pm$100 & 3300$\pm$400 & 760$\pm$90 \\
5 & 3s$^3$S$^0$$-$4p$^3$P & 4368.19 & 0.073$\pm$0.007 & 880$\pm$80 & 180$\pm$20 & $-$ & $-$ \\
10 & 3p$^5$P$-$4d$^5$D$^0$ & 6155.98 & 0.005: & 71: & 70: & $-$ & $-$ \\
20 & 3p$^3$P$-$5s$^3$S$^0$ & 7254.40 & 0.11$\pm$0.01 & 7300$\pm$600 & 2300$\pm$200 & $-$ & $-$ \\
21 & 3p$^3$P$-$4d$^3$D$^0$ & 7002.10 & 0.086$\pm$0.007 & 420$\pm$30 & 390$\pm$30 & $-$ & $-$ \\
22 & 3p$^3$P$-$6s$^3$S$^0$ & 6046.40 & 0.089$\pm$0.006 & 11500$\pm$800 & 5200$\pm$400 & $-$ & $-$ \\
23 & 3p$^3$P$-$5d$^3$D$^0$ & 5958.39 & 0.038$\pm$0.005 & 320$\pm$40 & 310$\pm$40 & $-$ & $-$ \\
24 & 3p$^3$P$-$7s$^3$S$^0$ & 5554.83 & 0.025$\pm$0.004 & $-$ & 3900$\pm$700 & $-$ & $-$ \\
25 & 3p$^3$P$-$6d$^3$D$^0$ & 5512.77 & 0.024$\pm$0.004 & 340$\pm$60 & 330$\pm$60 & $-$ & $-$ \\
26 & 3p$^3$P$-$8s$^3$S$^0$ & 5298.89 & 0.028$\pm$0.005 & $-$ & 11000$\pm$2000 & $-$ & $-$ \\
27 & 3p$^3$P$-$7d$^3$D$^0$ & 5274.97 & 0.011$\pm$0.003 & $-$ & 250$\pm$80 & $-$ & $-$ \\
\noalign{\vskip3pt} \noalign{\hrule} \noalign{\vskip3pt}
\multicolumn{3}{c}{Adopted} & & \multicolumn{4}{c}{\bf 14$\pm$4} \\
\noalign{\vskip3pt} \noalign{\hrule}
\end{tabular}
\begin{description}
\item[$^{\rm a}$] Effective recombination coefficients by Escalante \& Victor (1992).
\item[$^{\rm b}$] Effective recombination coefficients by P\'equignot et al. (1991).
\item[$^{\rm c}$] Blend with sky emission line.
\end{description}
\end{minipage}
\end{table*}

We have identified and measured a large number of {\oii} lines in our spectra. The largest collection 
of this kind of lines ever identified in an {\hii} region. In our inventory, there are lines coming from 
transitions between both possible kinds of levels: doublets and quartets. \citet{gra76} demonstrated the dominance of 
recombination in the excitation mechanism of the {\oii} spectrum. We have also measured 
several lines coming from 4$f-$3$d$ transitions and these lines cannot be excited by fluorescence from the 
2$p^3$ $^4S^0$ ground level. We have used effective recombination coefficients from \citet{sto94} for 3$s$-3$p$ 
and 3$p$-3$d$ transitions (assuming LS-coupling), and from \citet{liu95} for 3$p$-3$d$ and 3$d$-4$f$ transitions 
(assuming intermediate coupling). We used the dielectronic recombination coefficients of \citet{nus84} for 
multiplets 15, 16 and 36. The final adopted value of the  {\opp}/{\hydp} ratio has been obtained from the weighted 
mean of the sum values of those less case-dependent multiplets: number 1, 2 and 10 and all the 4$f-$3$d$ 
transitions. Our {\opp} abundance coincides with that obtained by EPTE for the same zone of the Orion nebula. 
All the individual abundance values used to derive the adopted average are indicated 
in boldface in Table 10. 

\setcounter{table}{9}
\begin{table*}
\centering
\begin{minipage}{150mm}
\caption{O$^{++}$/H$^+$ ratios from Permitted Lines}
\begin{tabular}{ccccccccccc}
\noalign{\hrule} \noalign{\vskip3pt}
& & & & \multicolumn{7}{c}{O$^{++}$/H$^+$ ($\times$10$^{-5}$)}\\
& & & $I$($\lambda$)/$I$(H$\beta$) & \multicolumn{3}{c}{S94$^{\rm a}$} & \multicolumn{3}{c}{LSBC95$^{\rm b}$} & 
NS84$^{\rm c}$\\
Mult. & Transition & $\lambda_0$ & ($\times$10$^{-2}$)& A & B & C & A & B & C & $-$\\
\noalign{\vskip3pt} \noalign{\hrule} \noalign{\vskip3pt}
1& 3s$^4$P$-$3p$^4$D$^0$& 4638.86& 0.057$\pm$0.005& 58$\pm$5& 56$\pm$5& $-$& $-$& $-$& $-$& $-$\\
 && 4641.81& 0.102$\pm$0.005& 37$\pm$2& 36$\pm$2& $-$& $-$& $-$& $-$& $-$\\
 && 4649.13& 0.155$\pm$0.005& 32$\pm$1& 31$\pm$1& $-$& $-$& $-$& $-$& $-$\\
 && 4650.84& 0.052$\pm$0.005& 54$\pm$5& 52$\pm$5& $-$& $-$& $-$& $-$& $-$\\
 && 4661.63& 0.068$\pm$0.005& 56$\pm$4& 54$\pm$4& $-$& $-$& $-$& $-$& $-$\\
 && 4673.73& 0.011$\pm$0.003& 70$\pm$20& 70$\pm$20& $-$& $-$& $-$& $-$& $-$\\
 && 4676.24& 0.035$\pm$0.005& 39$\pm$5& 37$\pm$5& $-$& $-$& $-$& $-$& $-$\\
 && 4696.36& 0.004:& 45:& 44:& $-$& $-$& $-$& $-$& $-$\\
 && Sum& 0.49$\pm$0.01& 40$\pm$1& {\bf 39$\pm$1}& $-$& $-$& $-$& $-$& $-$\\
2& 3s$^4$P$-$3p$^4$P$^0$& 4317.14& 0.044$\pm$0.005& 90$\pm$10& 61$\pm$7& $-$& $-$& $-$& $-$& $-$\\
 && 4319.63& 0.025$\pm$0.005& 49$\pm$9& 35$\pm$6& $-$& $-$& $-$& $-$& $-$\\
 && 4349.43& 0.065$\pm$0.006& 48$\pm$4& 34$\pm$3& $-$& $-$& $-$& $-$& $-$\\
 && 4366.89& 0.048$\pm$0.005& 78$\pm$9& 55$\pm$6& $-$& $-$& $-$& $-$& $-$\\
 && Sum& 0.23$\pm$0.01& 60$\pm$3& {\bf 43$\pm$2}& $-$& $-$& $-$& $-$& $-$\\
3& 3s$^4$P$-$3p$^4$S$^0$& 3712.74& 0.035:& 600:& 100:& $-$& $-$& $-$& $-$& $-$\\
 && 3749.48& 0.12$\pm$0.02& 600$\pm$100& 110$\pm$20& $-$& $-$& $-$& $-$& $-$\\
 && Sum& 0.22$\pm$0.02& 620$\pm$60& 110$\pm$10& $-$& $-$& $-$& $-$& $-$\\
4& 3s$^2$P$-$3p$^2$S$^0$& 6721.39& 0.006:& 100:& $-$& 80:& $-$& $-$& $-$& $-$\\
5& 3s$^2$P$-$3p$^2$D$^0$& 4414.90& 0.036$\pm$0.006& 70$\pm$10& $-$& 11$\pm$2& $-$& $-$& $-$& $-$\\
 && 4416.97& 0.028$\pm$0.004& 100$\pm$20& $-$& 16$\pm$3& $-$& $-$& $-$& $-$\\
 && Sum& 0.68$\pm$0.07& 82$\pm$8& $-$& 13$\pm$1& $-$& $-$& $-$& $-$\\
6& 3s$^2$P$-$3p$^2$P$^0$& 3973.24& 0.020$\pm$0.007& 80$\pm$30& $-$& 60$\pm$20& $-$& $-$& $-$& $-$\\
10& 3p$^4$D$^0$$-$3d$^4$F& 4069.62& 0.086$\pm$0.007& 34$\pm$3& $-$& $-$& 34$\pm$3& $-$& $-$& $-$\\
  && 4072.15& 0.067$\pm$0.006& 28$\pm$3& $-$& $-$& 28$\pm$3& $-$& $-$& $-$\\
  && 4075.86& 0.079$\pm$0.006& 23$\pm$2& $-$& $-$& 23$\pm$2& $-$& $-$& $-$\\
  && 4078.84& 0.011:& 20:& $-$& $-$& 28:& $-$& $-$& $-$\\
  && 4085.11& 0.013$\pm$0.004& 29$\pm$9& $-$& $-$& 26$\pm$8& $-$& $-$& $-$\\
  && 4092.93& 0.01:& 31:& $-$& $-$& 25:& $-$& $-$& $-$\\
  && Sum& 0.27$\pm$0.01& {\bf 27$\pm$1}& $-$& $-$& {\bf 27$\pm$1}& $-$& $-$& $-$\\
11& 3p$^4$D$^0$$-$3d$^4$P& 3864.12& 0.027:& 8000:& $-$& $-$& 11000:& 650:& 600:& $-$\\
12& 3p$^4$D$^0$$-$3d$^4$D& 3882.19& 0.021:& 34:& 33:& $-$& 63:& 61:& 33:& $-$\\
15& 3s$^2$D$-$3p$^2$F$^0$& 4590.97& 0.025$\pm$0.004& $-$& $-$& $-$& $-$& $-$& $-$& 160$\pm$30\\
  && 4595.95& 0.020$\pm$0.004& $-$& $-$& $-$& $-$& $-$& $-$& 150$\pm$30\\
  && Sum& 0.045$\pm$0.05& $-$& $-$& $-$& $-$& $-$& $-$& 150$\pm$20\\
16& 3s$^2$D$-$3p$^2$D$^0$& 4351.27& 0.008:& $-$& $-$& $-$& $-$& $-$& $-$& 50:\\
19& 3p$^4$P$^0$$-$3d$^4$P& 4121.46& 0.041$\pm$0.005& 3400$\pm$400& 130$\pm$17& $-$& 2600$\pm$300& 
150$\pm$20& 140$\pm$20& $-$\\
  && 4129.32& 0.008:& 4200:& 160:& $-$& 2000:& 120:& 110:& $-$\\
  && 4132.80& 0.033$\pm$0.005& 1500$\pm$200& 58$\pm$9& $-$& 1200$\pm$200& 
60$\pm$9& 56$\pm$8& $-$\\
  && 4153.30& 0.076$\pm$0.006& 2500$\pm$200& 96$\pm$8& $-$& 2200$\pm$200& 97$\pm$8& 91$\pm$7& $-$\\
  && Sum$^{\rm d}$& 0.190$\pm$0.01& 2400$\pm$100& 91$\pm$5& $-$& $-$& $-$& $-$& $-$\\
  && Sum$^{\rm e}$& 0.200$\pm$0.01& $-$& $-$& $-$& 1900$\pm$100& 94$\pm$5& 88$\pm$4& $-$\\
20& 3p$^4$P$^0$$-$3d$^4$D& 4104.99& 0.024$\pm$0.005& 25$\pm$5& 25$\pm$5& $-$& 400$\pm$80& 90$\pm$20& 
60$\pm$10& $-$\\
  && 4110.79& 0.024$\pm$0.005& 320$\pm$60& 310$\pm$60& $-$& 700$\pm$100& 100$\pm$20& 90$\pm$20& $-$\\
  && 4119.22& 0.031$\pm$0.005& 17$\pm$3& 17$\pm$3& $-$& 36$\pm$6& 35$\pm$6& 19$\pm$3& $-$\\
  && Sum$^{\rm d}$& 0.13$\pm$0.01& 28$\pm$2& 27$\pm$2& $-$& $-$& $-$& $-$& $-$\\
  && Sum$^{\rm f}$& 0.088$\pm$0.007& $-$& $-$& $-$& 77$\pm$6& $-$& $-$& $-$\\
  && Sum$^{\rm g}$& 0.102$\pm$0.008& $-$& $-$& $-$& $-$& 46$\pm$4& $-$& $-$\\ 
  && Sum$^{\rm h}$& 0.107$\pm$0.008& $-$& $-$& $-$& $-$& $-$& 28$\pm$2& $-$\\
25& 3p$^2$D$^0$$-$3d$^2$F& 4699.22& 0.010$\pm$0.003& 140$\pm$50& 7$\pm$2& $-$& 150$\pm$50& 130$\pm$40& 14$\pm$4& $-$\\
  && 4705.35& 0.018$\pm$0.004& 180$\pm$40& 9$\pm$2& $-$& 160$\pm$40& 160$\pm$30& 9$\pm$2& $-$\\
  && Sum& 0.028$\pm$0.003& 170$\pm$20& 8$\pm$1& $-$& 160$\pm$20& 150$\pm$20& 10$\pm$1& $-$\\
28& 3p$^4$S$^0$$-$3d$^4$P& 4890.86& 0.022$\pm$0.004& $-$& $-$& $-$& 3300$\pm$600& 190$\pm$40& 180$\pm$30& $-$\\
33& 3p$^2$P$^0$$-$3d$^2$D& 4943.00& 0.01:& 250:& 170:& $-$& 220:& 220:& 150:& $-$\\
36& 3p$^2$F$^0$$-$3d$^2$G& 4185.45& 0.021$\pm$0.004& $-$& $-$& $-$& $-$& $-$& $-$& 90$\pm$20\\
  && 4189.79& 0.025$\pm$0.005& $-$& $-$& $-$& $-$& $-$& $-$& 80$\pm$10\\
  && Sum& 0.046$\pm$0.005& $-$& $-$& $-$& $-$& $-$& $-$& 83$\pm$8\\
\noalign{\vskip3pt} \noalign{\hrule} 
\end{tabular}
\end{minipage}
\end{table*}

\setcounter{table}{9}
\begin{table*}
\centering
\begin{minipage}{150mm}
\caption{{\it --continued}}
\begin{tabular}{ccccccccccc}
\noalign{\hrule} \noalign{\vskip3pt}
& & & & \multicolumn{7}{c}{O$^{++}$/H$^+$ ($\times$10$^{-5}$)}\\
& & & $I$($\lambda$)/$I$(H$\beta$) & \multicolumn{3}{c}{S94$^{\rm a}$} & \multicolumn{3}{c}{LSBC95$^{\rm b}$} & 
NS84$^{\rm c}$\\
Mult. & Transition & $\lambda_0$ & ($\times$10$^{-2}$)& A & B & C & A & B & C & $-$\\
\noalign{\vskip3pt} \noalign{\hrule} \noalign{\vskip3pt}
3d-4f& 3d$^4$F$-$4fG$^2$[4]$^0$& 4083.90& 0.010$\pm$0.004& $-$& $-$& $-$& 30$\pm$10& $-$& $-$& $-$\\
     & 3d$^4$F$-$4fG$^2$[3]$^0$& 4087.15& 0.013$\pm$0.004& $-$& $-$& $-$& 40$\pm$10& $-$& $-$& $-$\\
     & 3d$^4$F$-$4fG$^2$[5]$^0$& 4089.29& 0.025$\pm$0.005& $-$& $-$& $-$& 22$\pm$4& $-$& $-$& $-$\\
     & 3d$^4$F$-$4fG$^2$[3]$^0$& 4095.64& 0.007:& $-$& $-$& $-$& 31:& $-$& $-$& $-$\\
     & 3d$^4$F$-$4fD$^2$[3]$^0$& 4107.09& 0.006:& $-$& $-$& $-$& 46:& $-$& $-$& $-$\\
     & 3d$^4$F$-$4fF$^2$[4]$^0$& 4062.94& 0.006:& $-$& $-$& $-$& 42:& $-$& $-$& $-$\\
     & 3d$^4$P$-$4fD$^2$[2]$^0$& 4307.23& 0.007:& $-$& $-$& $-$& 58:& $-$& $-$& $-$\\
     & 3d$^4$D$-$4fG$^2$[4]$^0$& 4332.69& 0.020$\pm$0.004& $-$& $-$& $-$& 180$\pm$40& $-$& $-$& $-$\\
     & 3d$^4$D$-$4fF$^2$[4]$^0$& 4275.55& 0.017$\pm$0.004& $-$& $-$& $-$& 27$\pm$6& $-$& $-$& $-$\\
     & 3d$^2$D$-$4fF$^2$[4]$^0$& 4609.44& 0.013$\pm$0.004& $-$& $-$& $-$& 27$\pm$7& $-$& $-$& $-$\\
     & 3d$^2$D$-$4fF$^2$[3]$^0$& 4602.11& 0.005:& $-$& $-$& $-$& 26:& $-$& $-$& $-$\\
     && Sum& 0.11$\pm$0.01& $-$& $-$& $-$& {\bf 30$\pm$3}& $-$& $-$& $-$\\
\noalign{\vskip3pt} \noalign{\hrule} \noalign{\vskip3pt}
\multicolumn{3}{c}{Adopted} & & \multicolumn{7}{c}{\bf 37$\pm$1} \\
\noalign{\vskip3pt} \noalign{\hrule} \noalign{\vskip3pt}
\end{tabular}
\begin{description}
\item[$^{\rm a}$] Effective recombination coefficients by Storey (1994).
\item[$^{\rm b}$] Effective recombination coefficients for intermediate coupling by Liu et al. (1995).
\item[$^{\rm c}$] Dielectronic recombination rates by Nussbaumer \& Storey (1984).
\item[$^{\rm d}$] Expected total intensity of the multiplet assuming LS coupling.
\item[$^{\rm e}$] Expected total intensity of the multiplet assuming intermediate coupling.
\item[$^{\rm f}$] Expected total intensity of the multiplet assuming intermediate coupling and case A.
\item[$^{\rm g}$] Expected total intensity of the multiplet assuming intermediate coupling and case B.
\item[$^{\rm h}$] Expected total intensity of the multiplet assuming intermediate coupling and case C.
\end{description}
\end{minipage}
\end{table*}

Several {\neii} lines are identified and measured in the blue spectral range covered with 
our data. These lines correspond to doublet, quartet and intercombination transitions. We have used the effective 
recombination coefficients computed by \citet{kis98} for deriving the {\nepp}/{\hydp} ratios shown in Table 11. 
We have used the quartet {\neii} lines to 
obtain the final adopted {\nepp} abundance (the weigthed average of the values obtained from the individual lines). 
These lines are case-independent and are very probably produced by 
pure recombination because the ground level has doublet configuration. In Figure 5 we show some of the quartet 
lines used to derive the {\nepp} abundance.  
This is the first time the {\nepp}/{\hydp} ratio is derived from recombination lines in the Orion nebula. 

\setcounter{table}{10}
\begin{table*}
\centering
\begin{minipage}{150mm}
\caption{Ne$^{++}$/H$^+$ ratios from Permitted Lines}
\begin{tabular}{cccccc}
\noalign{\hrule} \noalign{\vskip3pt}
& & & $I$($\lambda$)/$I$(H$\beta$) & \multicolumn{2}{c}{Ne$^{++}$/H$^+$ ($\times$10$^{-5}$)$^{\rm a}$} \\
Mult. & Transition & $\lambda_0$ & ($\times$10$^{-2}$) & A & B \\
\noalign{\vskip3pt} \noalign{\hrule} \noalign{\vskip3pt}
1 & 3s$^4$P$-$3p$^4$P$^0$ & 3694.22 & 0.04$\pm$0.01 & {\bf 12$\pm$4} & $-$ \\
2 & 3s$^4$P$-$3p$^4$D$^0$ & 3334.87 & 0.09$\pm$0.02 & {\bf 14$\pm$3} & $-$ \\
7 & 3s$^2$P$-$3p$^2$P$^0$ & 3323.75 & 0.06$\pm$0.02 & 20$\pm$7 & $-$ \\
19 & 3p$^2$D$^0$$-$3d$^4$F & 3388.46 & 0.03: & 10: & 9: \\
39 & 3p$^2$P$^0$$-$3d$^4$D & 3829.77 & 0.02: & 250: & 15 : \\
57 & 3d$^4$F$-$4f$^4$G$^0$ & 4391.94 & 0.014$\pm$0.004 & 4$\pm$1 & $-$ \\
& & 4409.30 & 0.009$\pm$0.003 & 4$\pm$1 & $-$ \\
&& Sum& 0.023$\pm$0.005 & {\bf 4$\pm$1} & $-$ \\
\noalign{\vskip3pt} \noalign{\hrule} \noalign{\vskip3pt}
\multicolumn{3}{c}{Adopted} & & \multicolumn{2}{c}{\bf 9$\pm$2} \\
\noalign{\vskip3pt} \noalign{\hrule} \noalign{\vskip3pt}
\end{tabular}
\begin{description}
\item[$^{\rm a}$] Effective recombination coefficients by Kisielius et al. (1998).
\end{description}
\end{minipage}
\end{table*}

\section{Ionic abundances from CELs and RLs and temperature variations}

Ionic abundances derived from CELs and RLs are systematically different in many ionized nebulae 
(e. g. Liu 2002, 2003; Esteban 2002, Torres-Peimbert \& Peimbert 2003). In fact, {\opp}/{\hydp} 
ratios obtained from {\oii} lines are between 0.1 to 0.3 dex larger than those obtained from [\oiii] 
lines in the few Galactic and extragalactic {\hii} regions where both kinds of lines have been 
observed (EPTE; Esteban et al. 1999a, b, 2003; Peimbert 2003; Tsamis et al. 2003). A similar  
situation has been found in the case of {\cpp}/{\hydp} and {\op}/{\hydp} ratios. In Table 12 we 
compare the different ionic abundances we have obtained from CELs and RLs of the same ions. 
The RLs abundances are the "Adopted" ones given in Tables 6 to 11. In the case of the 
{\cpp}/{\hydp} ratio obtained from CELs, we have taken the average of the values corresponding 
to slit positions 5 and 7 of \citet{wal92}. As it can be seen in Table 12, all the ionic 
abundances obtained from RLs are larger than the values derived from CELs.

\setcounter{table}{11}
\begin{table}
\centering
\begin{minipage}{150mm}
\caption{Abundance discrepancies and $t^2$ parameter}
\begin{tabular}{lccc}
\noalign{\hrule} \noalign{\vskip3pt}
& \multicolumn{2}{c}{12+log(X$^{\rm m}$/H$^+$)} &  \\
& CELs & RLs & $t^2$ \\
\noalign{\vskip3pt} \noalign{\hrule} \noalign{\vskip3pt}
O$^+$ & 7.76$\pm$0.15 & 8.15$\pm$0.13 & 0.052$\pm$0.029 \\
O$^{++}$ & 8.43$\pm$0.01 & 8.57$\pm$0.01 & 0.020$\pm$0.002 \\
C$^{++}$ & 7.94$\pm$0.15$^{\rm a}$ & 8.34$\pm$0.02 & 0.039$\pm$0.011 \\
Ne$^{++}$ & 7.69$\pm$0.07 & 7.95$\pm$0.07 & 0.032$\pm$0.014 \\
He$^+$ & ... & ... & 0.022$\pm$0.002 \\
$T$(Bac)/$T$(OII+OIII) & ... & ... & 0.018$\pm$0.018 \\
$T$(Pac)/$T$(OII+OIII) & ... & ... & 0.013$^{+0.033}_{-0.013}$ \\
Adopted & ... & ... & 0.022$\pm$0.002 \\
\noalign{\vskip3pt} \noalign{\hrule} \noalign{\vskip3pt}
\end{tabular}
\begin{description}
\item[$^{\rm a}$] Abundance taken from Walter et al. (1992)
\end{description}
\end{minipage}
\end{table}

\citet{tor80} proposed that the abundance discrepancy between calculations based 
on CELs and RLs may be produced by the presence of spatial fluctuations of the electron temperature 
in the nebulae, parametrized by $t^2$ \citep{pei67}. Assuming the validity of the temperature fluctuations 
paradigm, the comparison of the abundances determined from both kinds of lines for a given ion should 
provide an estimation of $t^2$. In Table 12 we include the $t^2$ values that produce the agreement 
between the abundance determinations obtained from CELs and RLs of {\op}, {\opp}, {\cpp} and {\nepp}. 
These calculations have been made following the formalism outlined by \citet{pei69}. As it can be seen 
in the table, the values of $t^2$ from the abundance discrepancies are --in general-- fairly similar 
taking into account the uncertainties. In Table 12 we also include the $t^2$ value obtained from the 
application of the maximum-likelihood method to the {\hep}/{\hp} ratios, obtained in Sect. 5. This 
value is in excellent agreement with that obtained for {\opp}. 
The comparison between electron temperatures obtained from intensity ratios of 
CELs and the Balmer or Paschen continua is an additional indicator of $t^2$. However, since $T_e$(Bac) and 
$T_e$(Pac) are representative of the whole nebula, the $T_e$ values obtained from CELs have to be 
considered only representative of the temperature of the zone where the ion producing the lines are located. 
Following Peimbert, Peimbert \& Luridiana (2002) and 
\citet{pei03}, we have compared $T_e$(Bac) and $T_e$(Pac) with the combination of $T$([OII]) and 
$T$([OIII]) considering a weight, $\gamma$, between the OII and OIII zones given by: 

\begin{equation}
\gamma = 
\frac {\int{N_e N({\rm O}^{++}) dV}}
{\int{N_e N({\rm O}^+) dV} + \int{N_e N({\rm O}^{++}) dV}}
.\end{equation}

Taking into account $\gamma$ $\approx$ 0.83 as representative for the center of the nebula (obtained from our 
derived abundances), we can obtain the average temperature $T$({\oii}+{\sc iii}) using equation A1 of \citet{pei02}, which 
gives: $T$({\oii}+{\sc iii}) = 8730$\pm$320 K. 
In Table 12, we include the values of $t^2$ obtained from the combination of $T$({\oii}+{\sc iii}) and 
$T$(Bac) and $T$(Pac). As we can see, the $t^2$ values obtained are rather consistent with the rest of determinations, 
especially with those obtained for {\opp} and {\hep}, the ones with the lowest uncertainties.  
However, the nominal $t^2$ values derived from the Balmer and 
Paschen discontinuities should be considered lower limits to the real ones. This is because 
we do not take into account the small Balmer and Paschen discontinuities that should be 
present in the nebular continua due to dust scattered light from the Trapezium stars 
\citep[see ][]{ode65}. It is beyond the scope of this paper to estimate the corrections to the 
temperatures due to this fact, but considering the large uncertainties of the 
$t^2$ determinations based on the discontinuities, its effect in the finally adopted weighted mean 
value of $t^2$ must be certainly negligible. 

We have calculated the weighted mean of the $t^2$ values given in Table 12  
to get a $t^2$ representative of the observed zone of the Orion nebula. The final adopted value is 
$t^2$ = 0.022$\pm$0.002. This result is consistent with those obtained by EPTE for the same 
zone: $t^2$ = 0.028$\pm$0.07, and their nearby Position 1: $t^2$ = 0.020$\pm$0.07. In addition, 
\citet{rub98} obtained an independent determination of $t^2$ = 0.032 from the comparison of the 
{\np}/{\op} ratios derived  
from optical and ultraviolet (UV) lines taken from the combination of {\it Hubble Space Telescope} ({\sc HST}) 
UV spectra of three zones of the Orion nebula. Finally, in a recent paper, \citet{ode03} have obtained a 
direct estimation of $t^2$ from the spatial changes in a high spatial resolution map 
(obtained from  
{\sc HST} images) columnar electron temperature of a region to the southwest of the Trapezium in the Orion 
nebula, very near our slit position. Their value is $t^2$ = 0.028$\pm$0.006. As it can be seen, 
it is very encouraging that different independent methods provide very 
consistent results, this suggest that temperature fluctuations are likely to be present in Orion nebula and that the true 
representative $t^2$ of its central parts should be between 0.020-0.030. 

\section{Total Abundances}

We have to adopt a set of ionization 
correction factors, ICFs, to correct for the unseen ionization stages  
in order to derive the total gaseous abundances of the different chemical elements. In our case, we adopt the 
ICF scheme used by EPTE for all the elements except Fe. For this element, we have determined the total abundance 
using two different ICFs. Firstly, we have considered our {\fepp} abundance and the ICF proposed by \citet{rod04}: 

\begin{equation}
\frac{N(\rm Fe)}{N(\rm H)} = \left[\frac{N(\rm O^{+})}{N(\rm O^{++})}\right]^{0.09}\times 
\frac{N(\rm Fe^{++})}{N(\rm O^{+})}\times \frac{N(\rm O)}{N(\rm H)}
.\end{equation}

Secondly, we have added our {\fepp} and {\feppp} abundances and include an ICF for the contribution of 
{\fep}. This contribution has been estimated from the observations of \citet{rod02}, who determine the 
{\fep} abundance from the {\ffeii} $\lambda$ 8617 \AA\ line. We have considered a 
{\fep}/{\fepp}=0.20, the average of the ratios obtained by \citet{rod02} for her four slit positions 
nearer the Trapezium cluster. The values of the ICFs assumed for the different chemical elements are included in Table 13. 

\setcounter{table}{12}
\begin{table}
\begin{minipage}{75mm}
\centering \caption{Adopted ICF values.}
\begin{tabular}{lcc}
\noalign{\hrule} \noalign{\vskip3pt}
Element & Unseen ion & Value \\
\noalign{\vskip3pt} \noalign{\hrule} \noalign{\vskip3pt}
He & He$^0$ & 1.12 \\
C & C$^{+}$ & 1.20 \\
N & N$^{++}$ & 5.68/5.90$^{\rm a}$ \\
Ne & Ne$^+$ & 1.60 \\
S & S$^{3+}$ & 1.10 \\
Ar & Ar$^+$ & 1.33 \\
Fe & Fe$^+$ & 1.07 \\
Fe & Fe$^+$, Fe$^{++}$ & 4.96/5.14$^{\rm a}$ \\
\noalign{\vskip3pt} \noalign{\hrule} \noalign{\vskip3pt}
\end{tabular}
\begin{description}
\item[$^{\rm a}$] Values for $t^2$=0.000/$t^2$=0.022
\end{description}
\end{minipage}
\end{table}

\setcounter{table}{13}
\begin{table}
\begin{minipage}{75mm}
\centering \caption{Total abundances$^{\rm a}$.}
\begin{tabular}{lcccc}
\noalign{\hrule} \noalign{\vskip3pt}
& \multicolumn{2}{c}{This Work} & \multicolumn{2}{c}{EPTE (Pos. 2)} \\
Element & $t^2$=0.000 & $t^2$=0.022$\pm$0.002 & $t^2$=0.000 & $t^2$=0.028 \\
\noalign{\vskip3pt} \noalign{\hrule} \noalign{\vskip3pt}
He & 10.991$\pm$0.003 & 10.988$\pm$0.003 & 11.00 & 10.99 \\
C$^{\rm b}$ & 8.42$\pm$0.02 & 8.42$\pm$0.02 & 8.37 & 8.37\\
N & 7.65$\pm$0.09 & 7.73$\pm$0.09 & 7.60 & 7.78 \\
O & 8.51$\pm$0.03 & 8.67$\pm$0.04 & 8.47 & 8.65 \\
O$^{\rm b}$ & 8.71$\pm$0.03 & 8.71$\pm$0.03 & ... & ... \\
O$^{\rm c}$ & 8.63$\pm$0.03 & 8.65$\pm$0.03 & ... & 8.68 \\
Ne & 7.78$\pm$0.07 & 8.05$\pm$0.07 & 7.69 & 7.89 \\
Ne$^{\rm b}$ & 8.16$\pm$0.09 & 8.16$\pm$0.09 & ... & ... \\
S & 7.06$\pm$0.04 & 7.22$\pm$0.04 & 7.01 & 7.24 \\
Cl & 5.33$\pm$0.04 & 5.46$\pm$0.04 & 5.17 & 5.37 \\
Ar & 6.50$\pm$0.05 & 6.62$\pm$0.05 & 6.53 & 6.86 \\
Fe$^{\rm d}$ & 6.07$\pm$0.08 & 6.23$\pm$0.08 & ... & ... \\
Fe$^{\rm e}$ & 5.86$\pm$0.10 & 5.99$\pm$0.10 & ... & ... \\
Fe$^{\rm f}$ & ... & ... & 6.27 & 6.34 \\
Fe$^{\rm g}$ & ... & ... & 6.01 & 6.07 \\
\noalign{\vskip3pt} \noalign{\hrule} \noalign{\vskip3pt}
\end{tabular}
\begin{description}
\item[$^{\rm a}$] In units of 12+log(X$^m$/H$^+$).
\item[$^{\rm b}$] Value derived from RLs.
\item[$^{\rm c}$] Value derived from {\oii} RLs and {\foii} CELs.
\item[$^{\rm d}$] Assuming $ICF$(Fe$^+$+Fe$^{3+}$).
\item[$^{\rm e}$] Assuming $ICF$(Fe$^+$).
\item[$^{\rm f}$] From Fe$^+$+Fe$^{++}$ and assuming $ICF$(Fe$^{3+}$).
\item[$^{\rm g}$] From Fe$^+$+Fe$^{++}$+Fe$^{3+}$. 
\end{description}
\end{minipage}
\end{table}

In Table 14 we show the total abundances obtained for our slit position of the Orion nebula. We include 
two different sets of abundances, one assuming no temperature fluctuations ($t^2$ = 0) and a second one using 
our final adopted value of $t^2$ = 0.022$\pm$0.002. In the table, we also compare with the abundances obtained 
by EPTE for their slit position 2, which coincides with our observed zone. We can see that the abundances 
are fairly similar in both set of data. Only Ne and Ar show differences larger than 0.1 dex. In the case of O 
we have included three sets of values: that obtained only from CELs, that obtained only from RLs and 
a last one that includes {\opp}/{\hydp} obtained from RLs and {\op}/{\hydp} obtained from CELs. We prefer 
this last determination because the {\op}/{\hydp} ratio determined from RLs is based on a single faint line 
located in a spectral zone with strong and numerous sky emission lines (see Sect. 7).
In the case of N, as it was commented in Sect. 7, we do not have considered the {\npp} abundance obtained 
from RLs because it gives abnormally large values of the final N/H ratio: 12+log(N/H) = 8.32$\pm$0.02 (for 
any of the two values of $t^2$ considered). This indicates that the observed {\nii} lines are not produced 
by pure recombination and an important contribution by fluorescence should be present. Finally, in the case of Fe, 
we find a ratio of about 1.9 in the two values of the Fe abundance given in Table 14. \citet{rod03} 
finds a similar result when comparing the Fe abundances of several objects. This author indicates that the most likely 
explanation of this discrepancy is that either the collision strengths of {\ffeiv} or the Fe ionization fractions 
predicted by ionization models (used for constructing Eq. 2) are unreliable. Unfortunately, we can not 
distinguish between these two possibilities. 

\section{Discussion}

\setcounter{table}{14}
\begin{table*}
\begin{minipage}{150mm}
\centering 
\caption{Chemical composition of different objects of the solar vicinity$^{\rm a}$.}
\begin{tabular}{lcccccc}
\noalign{\hrule} \noalign{\vskip3pt}
& Orion & & Young & & & \\
Element & gas+dust & Neutral ISM$^{\rm b}$ & F and G stars$^{\rm b}$ & B dwarfs$^{\rm c}$ & Sun$^{\rm d}$ & Orion$-$Sun \\
\noalign{\vskip3pt} \noalign{\hrule} \noalign{\vskip3pt}
He & 10.988$\pm$0.003 & ... & ... & ... & 10.98$\pm$0.02 & +0.008 \\
C & 8.52$\pm$0.02 & 8.15$\pm$0.06 & 8.55$\pm$0.10 & 8.25$\pm$0.08 & 8.41$\pm$0.05 & +0.11 \\
N & 7.73$\pm$0.09 & ... & ...& 7.81$\pm$0.09 & 7.80$\pm$0.05 & $-$0.07 \\
O & 8.73$\pm$0.03 & 8.50$\pm$0.02 & 8.65$\pm$0.15 & 8.68$\pm$0.06 & 8.66$\pm$0.05 & +0.07 \\
Ne & 8.05$\pm$0.07 & ... & ... & ... & 7.84$\pm$0.06 & +0.21 \\
S & 7.22$\pm$0.04 & ... & ... & ... & 7.20$\pm$0.08 & +0.02 \\
Cl & 5.46$\pm$0.04 & ... & ... & ... & 5.28$\pm$0.08 & +0.18 \\
Ar & 6.62$\pm$0.05 & ... & ... & ... & 6.18$\pm$0.08 & +0.44 \\
\noalign{\vskip3pt} \noalign{\hrule} \noalign{\vskip3pt}
\end{tabular}
\begin{description}
\item[$^{\rm a}$] In units of 12+log(X$^m$/H$^+$).
\item[$^{\rm b}$] Sofia \& Meyer (2001).
\item[$^{\rm c}$] Herrero (2003).
\item[$^{\rm d}$] Christensen-Dalsgaard (1998); Grevesse \& Sauval (1998); Asplund (2003); Asplund et al. (2004).
\end{description}
\end{minipage}
\end{table*}
 
The Orion nebula is traditionally considered the standard reference for the chemical composition
of the ionized gas in the solar neighborhood. Therefore, it is essential to have a confident 
determination of elemental abundances for this object. Until very recently it was thought that the Sun was a 
chemical anomaly because of its large abundances --specially O-- with respect to other 
nearby objects including the Orion nebula. In fact, at the beginning of the 90s the difference between the 
oxygen abundance of the Sun and the Orion nebula was about +0.4 dex (comparing the solar abundances of 
Grevesse \& Anders 1989 and those of the Orion nebula of Osterbrock et al. 1992). The recent corrections 
to the solar O abundance by \citet{asp04} have lowered it by a factor of 0.2 dex. On the other hand, 
our Orion nebula determinations based on RLs give also O/H ratios higher than the older ones by \citet{ost92}. 
However, for a correct comparison between solar and ionized gas abundances we have to correct for the fraction 
of heavy elements embedded in dust grains in the nebula. EPTE estimated that C and O abundances in Orion nebula should be 
depleted onto dust grains by factors of 0.10 dex and 0.08 dex, respectively. Adding this factors 
to the gaseous abundances we have appropriate values to comparing with the solar ones. 
In the cases of N, S and Cl, no dust correction is applied since they are not significantly depleted in the 
neutral interstellar medium \citep{sav96}. For He, Ne and Ar, no correction is necessary because they are 
noble gases. In Table 15 we compare our Orion nebula gas+dust abundances --corrected for depletion onto dust 
grains-- with those of the Sun, young F$-$G disk stars (ages $\leq$2 Gyr), nearby B dwarfs and gas-phase 
abundances of the local diffuse clouds. 
For the  Sun: He comes from \citet{chr98}; C and N from \citet{asp03}; O, Ne and Ar from \citet{asp04}, and S and Cl from 
\citet{gre98}. The data for F$-$G and B stars have been taken from the compilations by \citet{sof01} 
and \citet{her03}, respectively. The interstellar standard abundances of the nearby diffuse clouds have been taken from 
\citet{sof01}. 

The comparison of abundances given in Table 15 is very interesting. The O/H ratio of the Orion nebula is slightly higher 
but basically consistent within the uncertainties with the O abundance of young F$-$G stars, B dwarfs and the Sun.
This is a certainly remarkable result that does not longer support previous thoughts about the abnormally high 
chemical composition of the Sun with respect to other objects of the solar vicinity. 
In the case of C, the abundance is similar to that of F$-$G stars, somewhat higher than in the Sun and considerably 
higher than in B dwarfs. Nevertheless, the C abundance of B dwarfs could be erroneous because it could be affected 
by NLTE effects or problems with the C atomic model used as it has been pointed out by \citet{her03}. 
The N abundance of the Orion nebula is somewhat lower than in B dwarfs and the Sun, but consistent within the 
uncertainties. In the case of the other elements: Ne, S, Cl and Ar we can only compare with the Sun and their 
abundances are rather consistent except in the cases of Ne and Ar for which the differences are higher than 0.2 dex. 
Similar large differences for these elements are also reported in our data for the {\hii} region NGC~3576 \citep{gar04}. This 
indicates that those differences are not spurious but we cannot ascertain the exact reason for the discrepancy. 

The comparison with the abundances of nearby diffuse clouds is specially revealing. It is expected that 
C and O should be depleted onto dust grains in diffuse clouds (e.g. Jenkins 1987) and most probably in a larger amount 
than in ionized nebulae, where some dust destruction seems to operate (e.g. Rodr\'\i guez 1996). 
In this sense, the abundances obtained for diffuse clouds should be considered as lower limits of the expected 
ones in {\hii} regions. It is important to indicate that the comparison between the C and O abundances in diffuse clouds 
and those we obtain from CELs and 
assuming $t^2$=0.000 for the Orion nebula --8.02 and 8.51 for C and O, respectively-- do not give room for the 
expected dust destruction that should occur in ionized nebulae. The higher C and O abundances obtained from RLs --or 
from CELs assuming an appropriate $t^2$-- are more consistent with what is expected by the dust destruction scheme. 

The last column of Table 15 gives the difference between our Orion nebula abundances and the Solar ones. We find that 
most of the heavy elements give a positive difference, with an average value of about +0.09 dex (average of the
element values 
of Table 15 except He and Ar). This difference is 
in agreement with the estimations of the chemical evolution models by \citet{car03} and \citet{ake04} who found 
that the O/H ratio at the solar galactocentric distance has increased by 0.12 dex since the Sun was formed. 

Fe has not been included in Table 15 because large dust depletion factors are expected for this element in 
ionized nebulae. EPTE estimated a depletion of 1.37 dex comparing their gaseous Fe/H ratio with that of 7.48$\pm$0.15 
derived from B stars of the Orion association by \citet{cun94}. If we consider this last value as representative 
of the gas+dust Fe abundance of the Orion nebula, we obtain depletion factors of 1.25 and 1.49 dex depending on 
the final ICF scheme adopted to obtain the gaseous Fe/H ratio.

\section{Conclusions}

We present echelle spectroscopy in the 3100-10400 \AA\ range for the Orion nebula for a slit position 
coincident with previous observations of \citet{pei77} and EPTE. 
We have measured the intensity of 555 emission lines. This is the most complete list of emission lines 
ever obtained for this relevant object, and the largest collection of emission lines available for a 
Galactic or extragalactic {\hii} region. 

We have derived the physical conditions of the nebula making use of many different line intensities and continuum ratios. 
The chemical abundances have been derived making use of collisionally excited lines for a large number of ions as well 
as recombination lines for {\hep}, {\cpp}, {\op}, {\opp} and {\nepp}. In the case of {\op} and {\nepp} this is the first 
time that their abundance is derived from recombination lines. We have determined {\cpp} and {\opp} abundances from 
several lines corresponding to $f-d$ transitions that have not been observed in previous works. The abundances obtained from 
recombination 
lines are always larger than those derived from collisionally excited lines for all the ions where both kinds of lines are 
measured. We obtain remarkably consistent independent estimations of the temperature fluctuation parameter derived  
from different methods, which adopted average value is $t^2$ = 0.022$\pm$0.002, similar to other estimates from the literature. 
This result strongly suggests that moderate temperature fluctuations are present in the Orion nebula.

The Orion nebula is a standard reference for the chemical composition of the ionized gas of the solar vicinity and,
therefore, it is important to have a confident set of abundances for this object in order to improve our 
knowledge of the chemical evolution of this particular zone of the Galaxy. We have compared the 
chemical composition of the nebula with that of the Sun and other representative objects, as the neutral diffuse 
ISM, young F and G stars and B dwarfs of the solar vicinity. The abundances of the heavy elements 
in the Orion nebula are only 
slightly higher --about 0.09 dex-- than the solar ones, a difference that can be 
explained by the chemical evolution of the solar vicinity since the Sun was formed. The recent corrections to the 
solar abundances and our new values of the gas+dust Orion nebula abundances seem finally to converge, washing out 
the long-standing problem of the apparently abnormal solar abundances.

\section*{Acknowledgments}

We would like to thank R. Kisielius and P. J. Storey for providing us with their latest calculations of effective 
recombination coefficients for Ne and D. P. Smits for providing us unpublished atomic calculations for He. 
CE and JG would like to thank the members of the Instituto de Astronom\'\i a, UNAM, for their always 
warm hospitality. This work has 
been partially funded by the Spanish Ministerio de Ciencia y Tecnolog\'\i a (MCyT) under project AYA2001-0436. 
MP received partial support from DGAPA UNAM (grant IN114601). MTR received partial support from FONDAP(15010003), 
a Guggenheim Fellowship and Fondecyt(1010404). MR acknowledges support from Mexican CONACYT project J37680-E.

\begin{figure}
\begin{center}
\epsfig{file=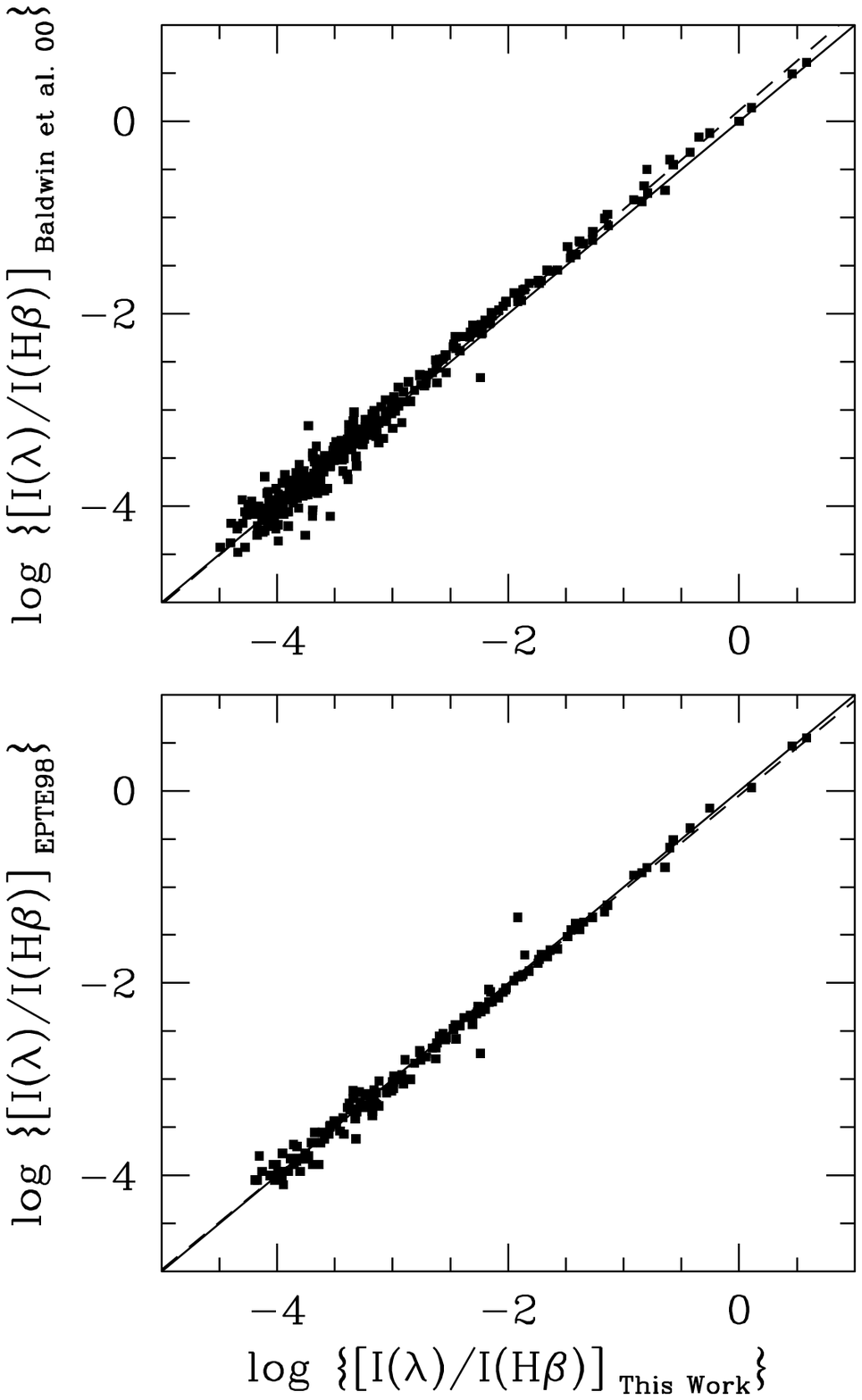, width=10. cm, clip=}
\caption{Comparison of line intensity ratios from this work with those of 
Baldwin et al. 2000 (top) and Esteban et al. 1998 (bottom). Continuous line represents the ideal 
relation with a slope of 1. Discontinuous line corresponds to the linear least-squares fit of the 
line ratios. }
\end{center}
\end{figure}

\begin{figure}
\begin{center}
\epsfig{file=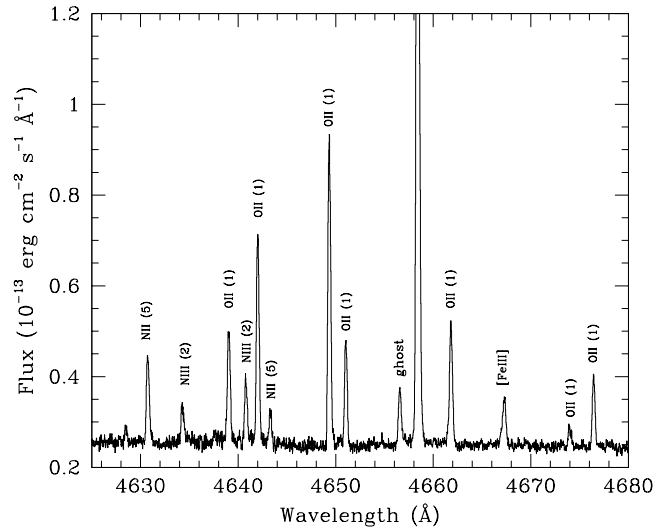, width=9. cm, clip=}
\caption{Section of the echelle spectrum showing all the individual emission 
lines of multiplet 1 of {\oii} (observed fluxes).}
\end{center}
\end{figure}

\begin{figure}
\begin{center}
\epsfig{file=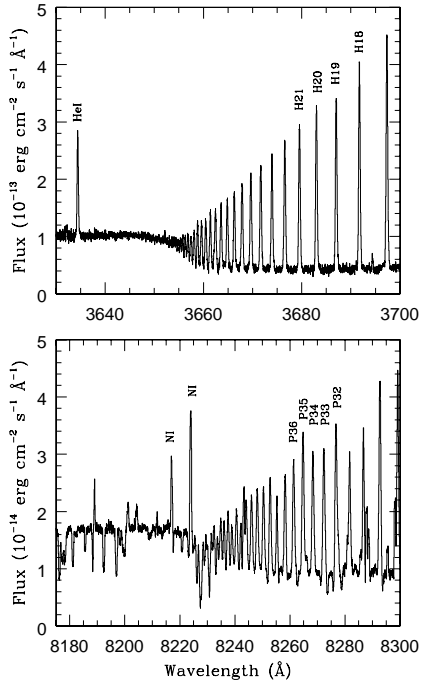, width=10. cm, clip=}
\caption{Section of the echelle spectrum showing the Balmer (top) and 
Paschen (bottom) discontinuities (observed fluxes).}
\end{center}
\end{figure}

\begin{figure}
\begin{center}
\epsfig{file=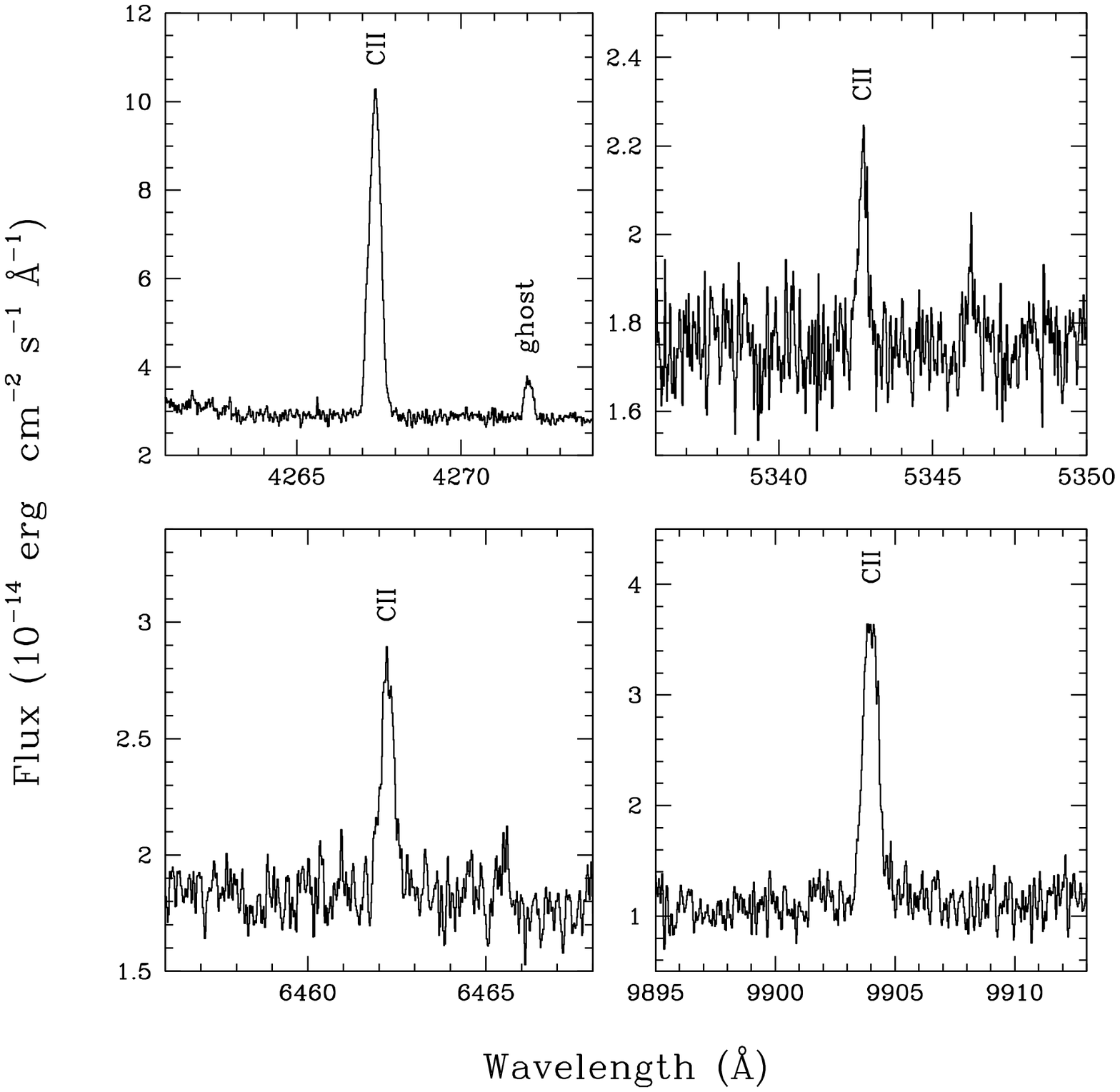, width=9. cm, clip=}
\caption{Section of the echelle spectrum showing some of the 
pure recombination {\cii} lines detected (observed fluxes).}
\end{center}
\end{figure}

\begin{figure}
\begin{center}
\epsfig{file=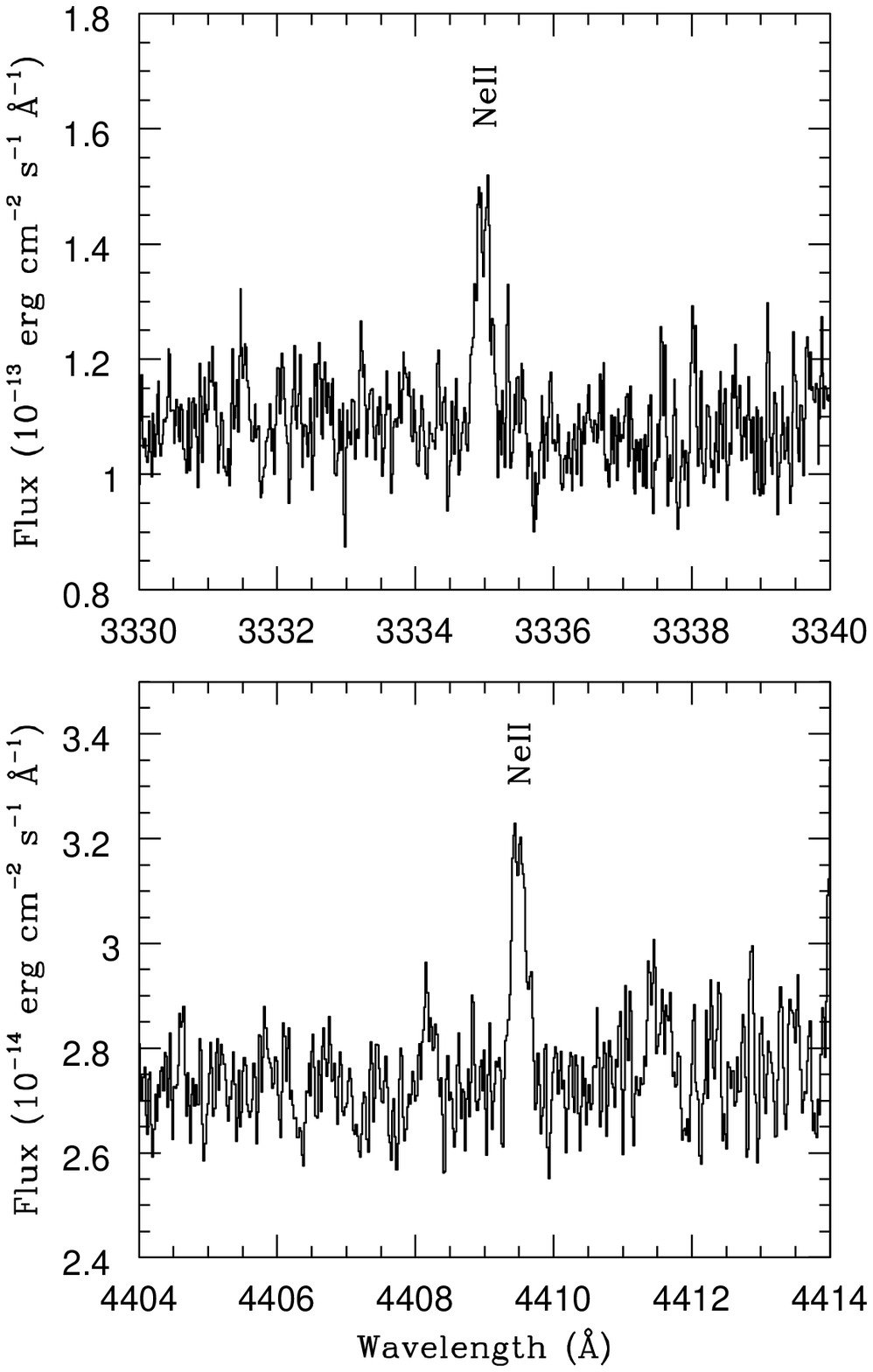, width=10. cm, clip=}
\caption{Section of the echelle spectrum showing some of the 
pure recombination {\neii} lines detected (observed fluxes).}
\end{center}
\end{figure}

\end{document}